\documentstyle[epsf,preprint,aps]{revtex}
\tightenlines
\def\bi{\begin{itemize}}
\def\ei{\end{itemize}}

\def\mrk#1{{\bf[[#1]]}}
\def\bcomment{\par\bgroup\advance\leftskip by0.5in\advance\leftmargin by0.5in\em}
\def\ecomment{\par\egroup\par}

\def\dddddd{\write16{%
============================================================================%
}}

\def\yy#1{\mrk{#1}\dddddd\write16{YY: #1}\dddddd}
\def\yy#1{\relax}

\begin{document}
\title{\bf TRANSIMS traffic flow characteristics}

\author{%
Kai Nagel$^{\dag}$, (505) 665-0921 phone, (505) 982-0565 fax, 
kai@lanl.gov email\\
~\\
Paula Stretz$^{\dag}$, (505) 665-6598 phone, (505) 665-7464 fax,
stretz@lanl.gov email\\
~\\
Martin Pieck$^{\dag}$, (505) 665-0086 phone, (505) 665-7464 fax,
martin@tsasa.lanl.gov email\\
~\\
Shannon Leckey$^{\dag}$, (505) 665-3733 phone, (505) 665-7464 fax,
shannon@tsasa.lanl.gov email\\
~\\
Rick Donnelly$^{\dag,*}$, (505) 665-3733 phone, (505) 665-7464 fax,
Rick.Donnelly@worldnet.att.net email\\
~\\
Christopher L.\ Barrett$^{\dag}$ (505) 665-0430 phone, (505) 665-7464 fax,
barrett@tsasa.lanl.gov email\\ 
~\\
${}^{\dag}$ Los Alamos National Laboratory,
MS 997, Los Alamos NM 87545, USA\\
~\\
${}^*$ Parsons Brinckerhoff, Inc., 5801 Osuna Road NE \# 220,
Albuquerque NM 87109, USA\\
~\\
PREPRINT \today
}
\maketitle
\setcounter{page}{0}
\begin{abstract}\noindent
Knowledge of fundamental traffic flow characteristics of traffic
simulation models is an essential requirement when using these models
for the planning, design, and operation of transportation systems.  In
this paper we discuss the following: a description of how features
relevant to traffic flow are currently under implementation in the
TRANSIMS microsimulation, a proposition for standardized traffic flow
tests for traffic simulation models, and the results of these tests
for two different versions of the TRANSIMS microsimulation.

keywords: traffic simulation, traffic flow, intersections
\end{abstract}

\vfill\eject
\setcounter{page}{0}
~\vfill\eject
\setcounter{page}{1}

\section{Introduction}
One could probably reach agreement that the traffic flow behavior of
traffic simulation models should be well documented.  Yet, in practice,
this turns out to be somewhat difficult.  Many traffic simulation models
are under continuous development, and the traffic flow dynamics documented
in a certain publication has probably been refined and extended until the
paper gets actually published.  

It makes thus sense to agree on a certain set of tests for traffic
flow dynamics which should always be run and documented together with
any ``real'' results.  In this paper, we propose such a suite of
traffic flow measurements.  We are well aware of the fact that some of
the results in this paper are arguably unrefined with respect to
reality.  Yet, as we stated above, we are continuously working on
improvements, and this publication represents both a snapshot of where
we currently stand and an argument for a standardized traffic flow
test suite for simulation models.  We hope that this publication will
both open the way for a constructive dialogue on which standardized
traffic flow tests should be run for traffic simulation models, and
which of the features of our traffic simulation models may need
improvement.

When designing a traffic microsimulation model, the first idea might
be to measure all aspects of human driving and put them in algorithmic
form into the computer.  Unfortunately, such attempts cause many
problems. The first is a data collection problem, because one can
certainly not measure ``all'' aspects of human driving and is thus
faced with the double sided problem that the necessary data collection
process is extremely costly and still selective.  Second, what if the
macroscopic flow properties of such a model are clearly wrong, for
example producing an hourly flow rate that is much to high or to low?
Since, in such a modeling approach, one does not know the connection
between the many parameters of the model and the emergent properties
(such as flow), one is left to random trial and error.

For that reason, the TRANSIMS (TRansportation ANalysis and SIMulation
System~\cite{TRANSIMS}) microsimulation starts with a {\it minimal\/}
approach.  A minimal set of driving rules is used to simulate traffic,
and this set of rules is only extended when it becomes clear that a
certain important aspect of traffic flow behavior cannot be included
with the current rule set.  Besides the conceptual clarity, this also
has the advantage that it is usually computationally fast -- minimal
models have few rules and thus run fast on computers.  This argument
also makes it clear that one wants to remain flexible with respect to
refinements of the model: If certain refinements are unnecessary with
respect to a certain question, one would want to switch them off both
for conceptual clarity and for computational speed.

The questions that TRANSIMS is currently designed for are transportation
{\em planning\/} questions.  The most important zeroth order result of a
transportation microsimulation should be the {\em delays}, since, once they
occur, they dominate travel times, and also hinder discharge of the
transportation system, thus leading to grid-lock.  Delays are caused by
congestion, and congestion is caused by demand being higher than capacity.
This implies that the first thing the TRANSIMS traffic microsimulation has
to get right are capacity constraints (and possibly their variance).
Capacity constraints are caused by a variety of effects:\bi

\item
Undisturbed roadways such as freeways have capacity constraints given
by the maximum of the flow-density diagram.

\item
Typical arterials have their capacity constraints given by traffic
lights.

\item
In the case of unprotected turning movements (yield, stop, ramps,
unprotected left, etc.), the capacity constraints are given as a
function of the traffic on the ``interfering lanes''.  For example,
the number of vehicles making an unprotected left turn depends on the
oncoming traffic.

\ei
Building a simulation which can be adjusted against all these diagrams
seems a hopeless task given the enormous amount of degrees of
freedom.  The TRANSIMS approach for that reason has been to {\em
generate\/} the correct behavior from a few much more basic
parameters.   The correct behavior with respect to the above criteria
can essentially be obtained by adjusting two parameters: (i) The value
of a certain asymmetric noise parameter in the acceleration determines
maximum flow on freeways and through traffic lights; (ii) the value of
the gap acceptance determines flow for unprotected movements.

The remainder of this paper will first describe the algorithms
TRANSIMS uses for the most important traffic movements, and then
describe the resulting flow characteristics.

\section{Rules}
\subsection{Single lane uni-directional traffic}
\begin{figure}
\centerline{%
\epsfxsize0.5\hsize
\epsfbox[0 0 542 152]{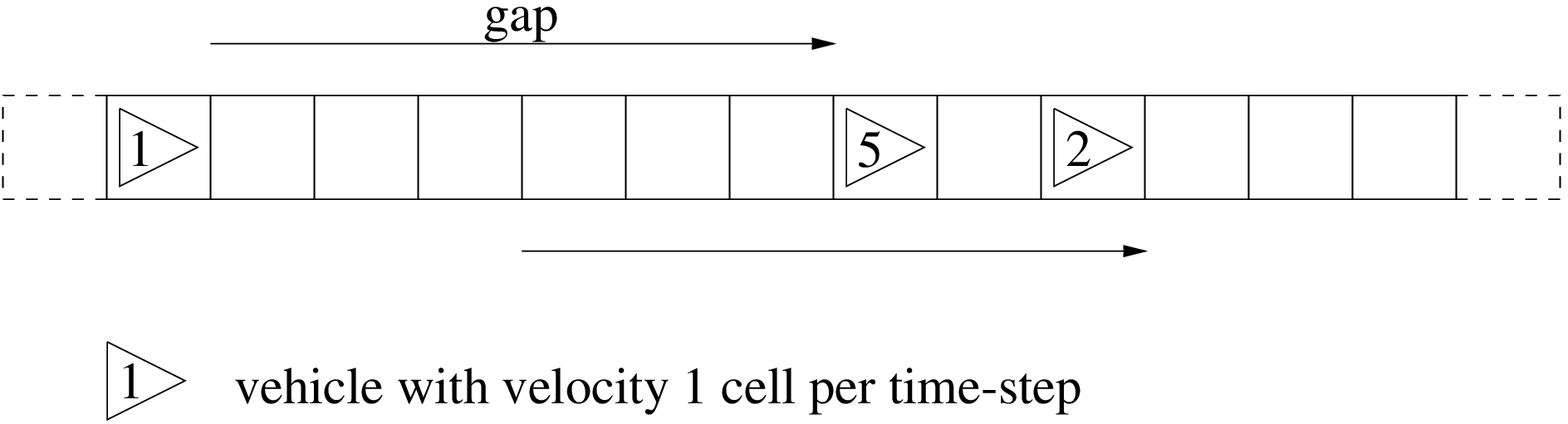} \hfill (a)
\hfill\hfill
\epsfxsize0.5\hsize
\epsfbox[0 0 758 411]{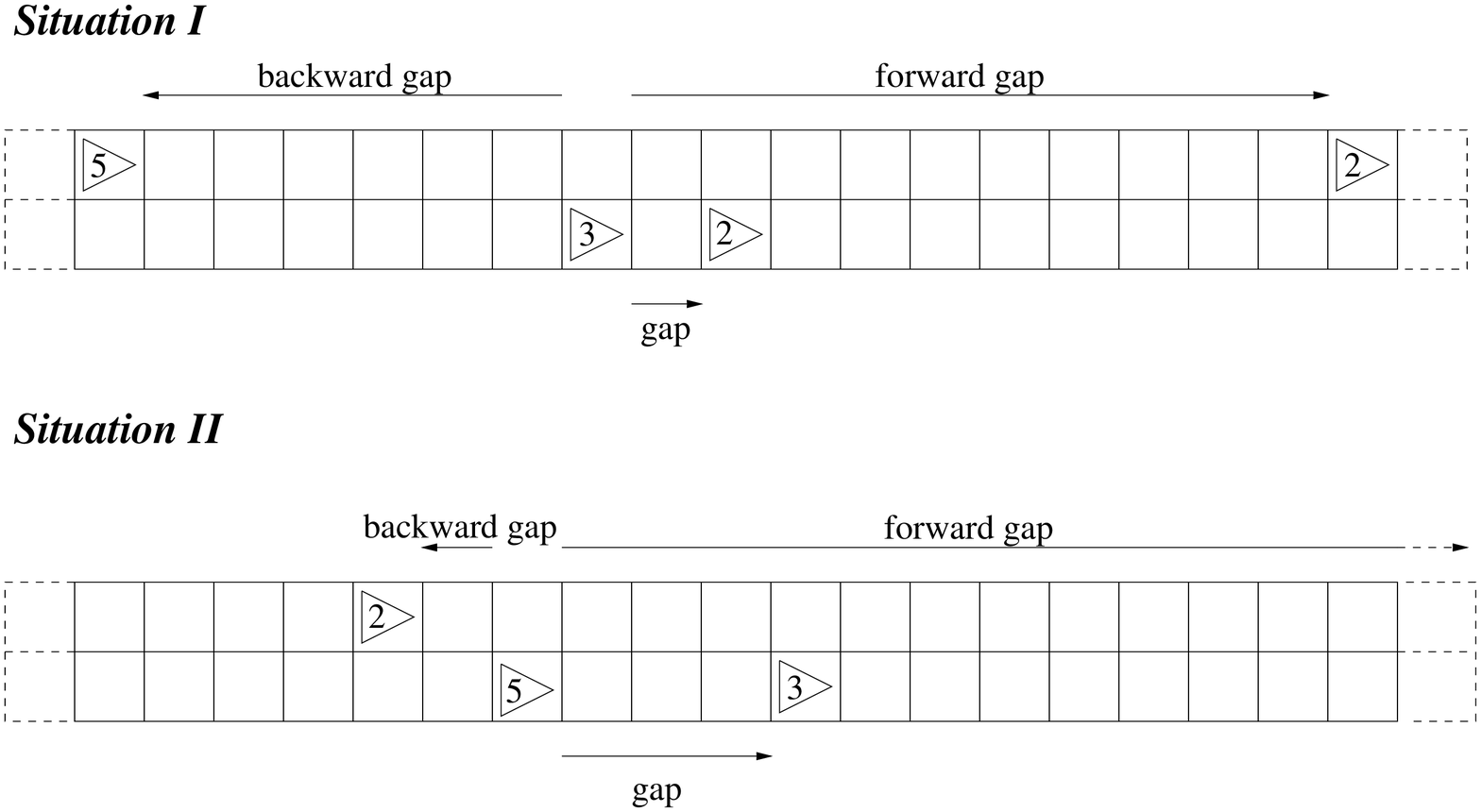} \hfill (b)
}

\vskip2cm

\centerline{%
\epsfxsize0.5\hsize
\epsfbox[0 0 758 288]{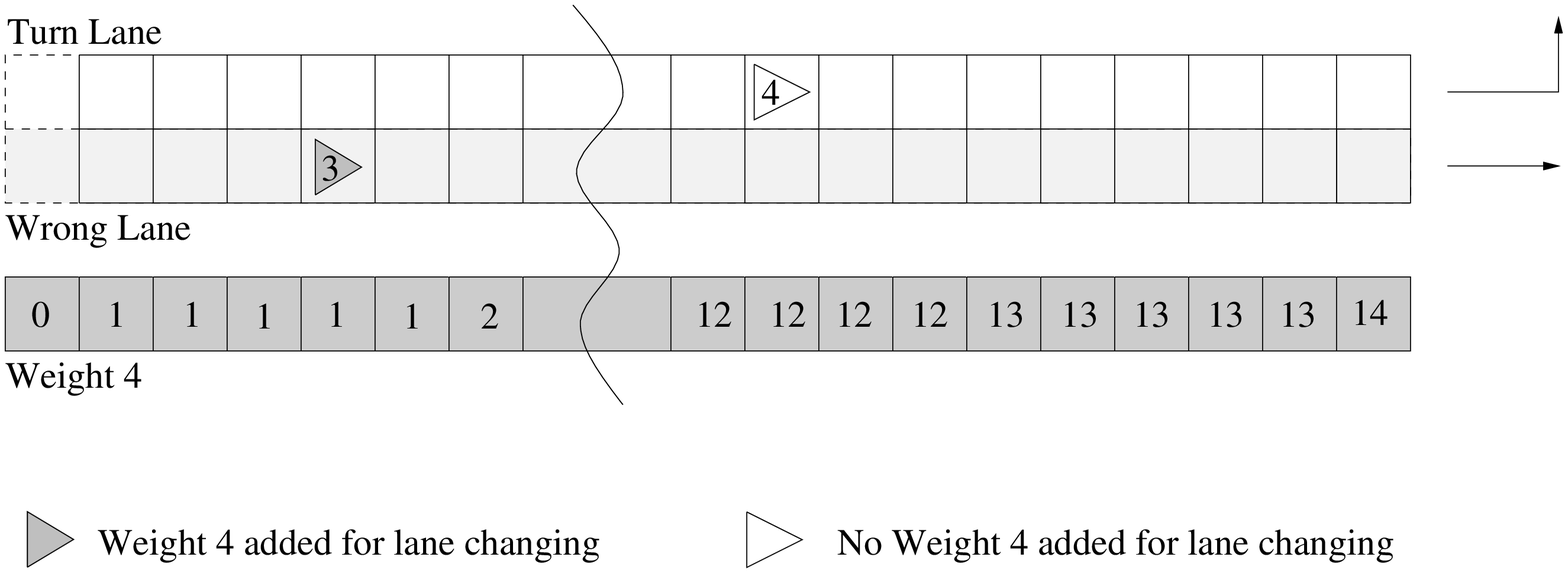} \hfill (c)
\hfill\hfill
\epsfxsize0.5\hsize
\epsfbox[0 0 758 419]{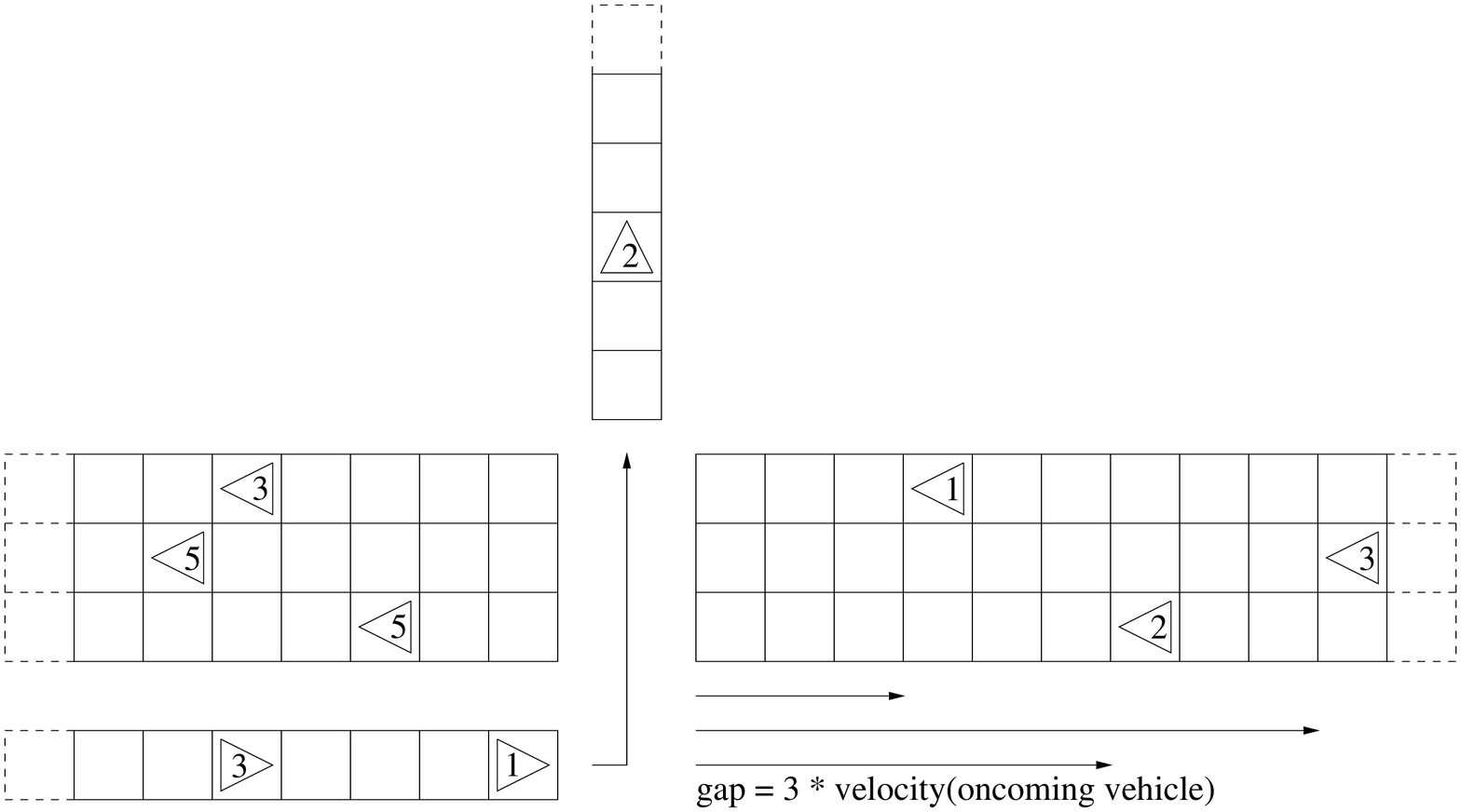} \hfill (d)
}

\vskip2cm

\epsfxsize0.7\hsize
\centerline{\hfill\epsfbox[0 0 686 290]{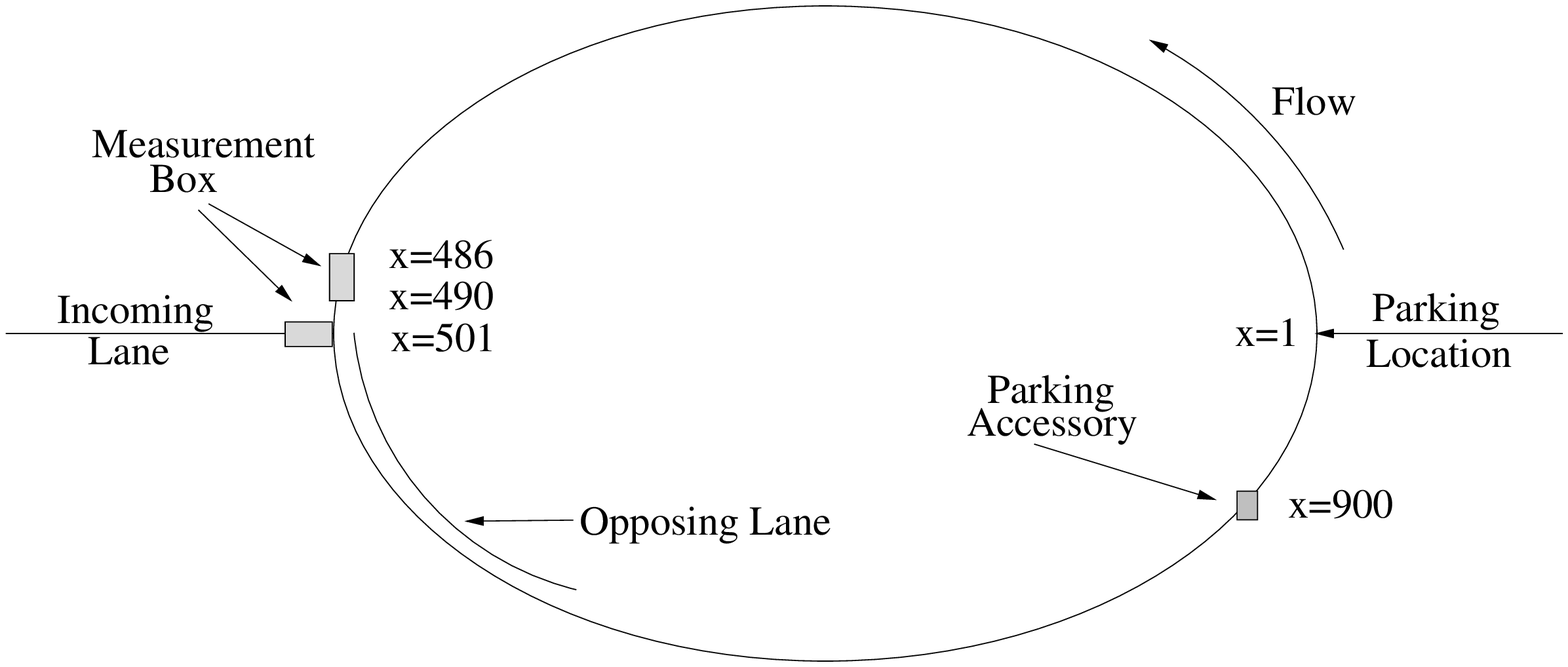} \hfill (e)}

\caption{\label{defs}%
{\em (a)} Definition of $gap$ and examples for one-lane update rules.
Traffic is moving to the right.  The leftmost vehicle accelerates to
velocity~2 with probability 0.8 and stays at velocity~1 with
probability 0.2.  The middle vehicle slows down to velocity~1 with
probability 0.8 and to velocity~0 with probability 0.2.  The right
most vehicle accelerates to velocity~3 with probability 0.8 and stays
at velocity~2 with probability 0.2.  Velocities are in ``cells per
time step''.  All vehicles are moved according to their velocities at
a later phase of the update.  
{\em (b)} Illustration of lane changing rules.  Traffic is moving to
the right; only lane changes to the left are considered.  Situation I:
The leftmost vehicle on the bottom lane will change to the left
because (i)~the forward gap on its own lane, 1, is smaller than its
velocity, 3; (ii)~the forward gap in the other lane, 10, is larger
than the gap on its own lane, 1; (iii)~the forward gap is large enough
compared to its own velocity: $weight2 = v - gap_f = 3 - 10 = -7 < 1 =
weight1$; (iv)~the backward gap is large enough: $weight3 = v_{max} -
gap_b = 5 - 6 = -1 < 1 = weight1$.  Situation II: The second vehicle
from the right on the right lane will not accept a lane change because
the gap backwards on the target lane is not sufficient.
{\em (c)} Value of $weight4$ when in wrong lane during the approach to
the intersection.
{\em (d)} Example of a left turn against oncoming traffic.  The turn
is accepted because on all three oncoming lanes, the gap is larger or
equal to three times the first oncoming vehicle's velocity.
{\em (e)} Test networks.
}
\end{figure}

Our traffic simulation is based on a cellular automata technique,
i.e., a road is composed of cells, and each cell can either be empty,
or occupied by exactly one
vehicle~\cite{Nagel:PhysComp,Nagel:Schreck}, see Fig.~\ref{defs}~(a).
Since movement has to be from one cell to another cell, velocities
have to be integer numbers between 0 and $v_{max}$, where the unit of
velocity is [cells per time-step].  It turns out that reasonable
values are~\cite{Nagel:Schreck,multi:res}:\bi

\item
length of a box $= 1 / \rho_{jam} =$ 7.5~m ($\rho_{jam} =$ density of
vehicles in a jam).

\item
time step $=$ 1~sec

\item
maximum velocity $=$ 5~boxes per time step $= 5 \cdot 7.5~\hbox{m} /
\hbox{sec} = 135 \hbox{km/h} \approx 85 \hbox{mph}$

\ei For other conditions, such as higher or lower speed limits, this
can be adapted.

Note that this approach implies a {\em coarse graining\/} of the spatial
and temporal resolution and therefore of the velocities.  A vehicle which
has a speed of, say, 4 in this model stands for a vehicle which has a speed
anywhere between $3.5 \cdot 7.5$ meters/sec $\approx$ 95~km/h (59~mph)
and $4.49999 \cdot 7.5$ meters/sec $\approx$ 121 km/h (75~mph).

Vehicles move only in one direction. For an arbitrary configuration
(velocity and position), one update of the traffic system consists of
two steps: a velocity update step consisting of three consecutive
rules, and a movement step according to the result of the velocity
update. The whole update is performed simultaneously for all
vehicles. The complete configuration at time step $t$ is stored and
the configuration at time step $t+1$ is computed from that ``old''
information.  Computationally we calculate in time step $t$ (with the
three rules) the new velocity of each car and write this newly
calculated velocity in the same site without moving the car (velocity
update). After that we move all cars according to their newly
calculated velocity (movement update).

\begin{enumerate}

\item (velocity update)

For all particles~$i$ simultaneously, do the following:

{\bf IF} ( $v_i \ge gap_i$ ) 

\qquad $\displaystyle v_i := \cases{ 
gap_i-1 & with probability $p_{noise}$ if possible\footnote{%
If this rule would ask for a negative velocity, then $v=0$ is chosen.
}\cr
gap_i & else \cr }$  (close following/braking)

{\bf ELSE IF} ( $v_i < v_{max}$ )

\qquad $\displaystyle v_i := \cases{
v_i & with probability $p_{noise}$ \cr
v_i + 1 & else \cr }$ (acceleration)

{\bf ELSE}  (i.e.\ ( $v_i = v_{max}$ AND $v_i < gap_i$ )

\qquad $\displaystyle v_i := \cases{
v_{max}-1 & with probability $p_{noise}$ \cr
v_{max} & else \cr }$ (free driving)

{\bf ENDIF}

\item (movement update)

Move all particles~$i$ to $x_i(t+1) = x_i(t) + v_i$.

\end{enumerate}
The index {\it i} denotes the position (an integer number) of a vehicle,
{\it v\/}({\it i\/}) its current velocity, {\it v\/}$_{\it max\/}$ its
maximum speed, {\it gap\/}({\it i\/}) the number of empty cells
ahead, and $p_{noise}$ is a randomization parameter.

The first velocity rule represents noisy car following or braking.  If
the vehicle ahead is too close, the vehicle itself attempts to adjusts
its velocity such that it would, in the next time-step, reach a
position just behind where the vehicle ahead is at the moment.  Yet,
with probability~$p_{noise}$, the vehicle is a bit slower than this.

The second velocity rule represents noisy acceleration.  Essentially,
the acceleration is linear (i.e.\ independent from current speed), but
with probability $p_{noise}$, no acceleration happens in the current
time step (maybe as a result of switching gears etc.).  Instead of an
acceleration sequence of $0 \to 1 \to 2 \to 3 \to \ldots$, a possible
acceleration sequence can now be $0 \to 0 \to 1 \to 2 \to 2 \to 2 \to 3 \to
\ldots$.

The last velocity rule represents free driving.  Instead of remaining
always at the same speed, such vehicles fluctuate between $v_{max}$
(with probability $1-p_{noise}$) and $v_{max}-1$ (with probability
$p_{noise}$).  Note that a vehicle which is set to $v_{max}-1$ will go
through the acceleration step next time, thus in the next time step
either staying at $v_{max}-1$ with probability $p_{noise}$ or getting
back to $v_{max}$.  Note that the resulting average speed of a freely
driving vehicle is thus $v_{max}-p_{noise}$.

This microsimulation is also fairly well understood from a theoretical
perspective; see~\cite{Nagel:flow:pre,Nagel:julich} for more
information.

\subsection{Lane changing for passing}
For multi-lane traffic, the model consists of parallel single lane
models with additional rules for lane changing.  Here we describe the
two lane model which can be modified to any kind of multi lane model.
Lane changing is modeled by an additional update step, which is added
before the velocity update.  The new sequence of steps is presented
below. Steps two and three are the same in the single lane model and
they are executed separately for each lane.
\begin{enumerate}
\item {\bf Lane changing decision}
\item {\bf Velocity update}
\item {\bf Vehicle movement}
\end{enumerate}  

According to this lane changing rule set the vehicles are only moving
sideways during the lane changing step; forwards movement is done in
the vehicle movement step.  One should, though, look at the combined
effect of the lane changing and the movement, and then vehicles will
usually have moved sideways {\em and\/} forwards.  The decision to
change lane is implemented as strictly parallel update, i.e. each
vehicle is making its decision based upon the configuration at the
beginning of the update.
\begin{itemize}
\item {\bf Lane changing decision for passing}
\begin{itemize}
\item {\bf IF} neighboring position $x_o(i)$ in other lane is vacant
\begin{itemize}
\item {\bf THEN }Calculate:
\begin{itemize}
\item {\it gap\/}({\it i\/}) Gap Forward in Current Lane,
\item ${\it gap_o\/}$({\it i\/}) Gap Forward in Other Lane,
\item ${\it gap_b\/}$({\it i\/}) Gap Backward in Other Lane,
\item {\bf IF (}{\it gap\/}({\it i\/}) $<$ {\it v\/}({\it
i\/}) AND ${\it gap_o\/}$({\it i\/}) $> gap(i)$ {\bf ~)}\\
$\;\;\;\;\;-\;\;\;${\bf THEN} $weight1 = 1$\\
$\;\;\;\;\;-\;\;\;${\bf ELSE} $weight1 = 0$
\item $weight2 = v(i) - gap_f(i)$
\item $weight3 = v_{max}(i) - gap_b(i)$.
\end{itemize}
\item {\bf IF} {\bf (} $weight1 > weight2$ {\bf )} AND {\bf (} $weight1
> weight3$ {\bf )}\footnote{%
Weights are used because of extensibility towards ``lane changing for
plan following''.  See below.
}
\begin{itemize}
\item {\bf THEN} mark vehicle for lane change\footnote{%
In the current version, the lane change is actually still rejected
with a probability of 0.01 even when all the rules are fulfilled.
This is in order to break the following artifact or variations of it:
Assume one lane is completely occupied and one is completely empty.
The above rule set will result in these vehicles just changing back
and forth between the lanes---the vehicles will never get smeared out
across the lanes.  See Ref.~\protect\cite{Rickert:etc:twolane} for
more details.
}
\end{itemize}
\end{itemize}
\end{itemize}  
\end{itemize}
The rules are working in the following way (see Fig.~\ref{defs}~(b)):
First we look at the neighboring position in the target lane.  If this
cell is vacant, we calculate the gap forward in the current lane
($gap$), the gap forward in the target lane ($gap_o$), and the gap
backward in the target lane ($gap_b$).  With these results we
calculate the $weight1$ to $weight3$ described above. Finally if the
weight comparisons render true the car will change to the new lane.
After executing the lane changing decision we calculate the new
velocity for all cars and move them according to this velocity.

For three or more lanes, a simultaneous implementation of the lane changing
decision can lead to collisions.  For example, in a three-lane road two
vehicles on the left and right lane could decide to go to the same spot in
the middle lane.  {From} an algorithmic point of view, this is possible
because the lane changing decision is based on the configuration on time
{\it t}; but it is also an entirely realistic situation.\footnote{%
In a deeper sense, the problem is caused by the fact that the
underlying decision making dynamics has a time scale which is smaller
than the time resolution of the simulation.  The simulation thus must
resolve the conflict by other means.
} To avoid collision we only allow lane changes in a certain direction
in each time step:
\begin{itemize}

\item {\bf IF} the time step is even

{\bf THEN} start procedure {\it lane changing decision}
to the {\it left} for cars on the middle and then on the right
lane

\item {\bf IF} the time step is odd

{\bf THEN} start procedure {\it lane changing decision}
to the {\it right} side for cars on the middle and then on the left
lane

\end{itemize}
Thus, left lane changes occur only on even time steps, right lane
changes occur only on odd time steps.  This behavior is collision
free.

\subsection{Lane changing for plan following}
Vehicles in TRANSIMS follow route plans, i.e.\ they know ahead of time
the sequence of links they intend to follow.  This means that, when
they approach an intersection, they need to get into the correct lanes
in order to make the intended turn.  For example, a vehicle which
intends, according to its route plan, to make a left turn at the next
intersection needs to get into one of the lanes which actually allow
a left turn.

This is achieved in TRANSIMS by supplementing the basic lane changing
rules with a bias towards the intended lanes.  This bias increases
with increasing urgency, i.e.\ with decreasing distance to the
intersection.  Technically, this is achieved by adding another weight
to the acceptance conditions for lane changing:\begin{itemize}

\item
{\bf IF} ($weight1 + weight4 > weight2$) AND ($weight1 + weight4 > weight3$)

{\bf THEN} change lane

\end{itemize}
$weight4$ is calculated according to
\[
weight4 = \max\left[ { d^* - d \over v_{max} } , 0 \right]
\]
for lane changes in the desired direction as long as the vehicle is
not in one of the correct lanes, cf.\ Fig.~\ref{defs}~(c).  $d$ is the
remaining distance to the intersection, $d^*$ is a parameter; both are
given in the unit of ``cells''.  $d^*$ is currently set to 70~cells,
i.e.\ approx.\ 500~m or 1/3~of a mile, throughout the simulation.  In
consequence, $weight4$ increases from zero to $d^*/v_{max}=14$ during
the approach to the intersection.  If $weight4 = 0$, then it does not
influence lane changing decision.  $weight4 = 1$ has the same effect
as a slower vehicle ahead on the same lane.  Further increases of
$weight4$ more and more override the security criterions that the
forward and the backward gap on the destination lane need to be large
enough.  $weight4 > v_{max}$ lets the vehicle make the lane change
even if only the neighboring cell on the destination lane is free.

\subsection{Unprotected turning movements}
\label{unprotected}
A necessary element of traffic simulations are unprotected turning
movements.  By this we mean that that for the movement the driver intends
to make, some other lanes have priority.  Examples are stop signs, yield
signs, on-ramps, unprotected left turns.

The general modeling principle for this in TRANSIMS is based on a gap
acceptance in the interfering lanes, see Fig.~\ref{defs}~(d).
Interfering lanes are the lanes which have priority; for example, for
a stop-controlled left turn onto a major road this would be all lanes
coming from the left plus the leftmost lane coming from the right.  In
order to accept the turn, there has to be a sufficient gap on each of
these lanes.

Note that ``gap divided by the velocity of the oncoming vehicle'' is the
oncoming vehicle's time headway, which is the typical measure used in the
Highway Capacity Manual~\cite{HCM}.  If one wants a time headway on an
interfering lane of at least 3~seconds, then a vehicle with a velocity of
4~cells/second would have to be at least 12~cells away from the
intersection.

The current TRANSIMS microsimulation uses a gap acceptance (gap
between intersection and nearest car to the intersection which is
approaching) of 3~times the oncoming vehicle's velocity, i.e.\ when
the gap on each interfering lane is larger than or equal to the first
vehicle on that lane, the move is accepted. For example, if the
oncoming vehicle has a speed of~3, at least~9 empty cells have to be
between the oncoming vehicle and the intersection.  A special case is
if the oncoming vehicle has the velocity zero, in which case no gap is
necessary.\footnote{%
The condition for the ``case study'' microsimulation of
TRANSIMS~\protect\cite{Beckman:etc:case-study,Nagel:Barrett:feedback}
was that a movement was accepted if, for all interfering lanes, the
gap was larger than $v_{max}$.  That means that for fast oncoming
traffic the acceptance was higher than in the newer version, but for
low speed oncoming traffic the acceptance rate was lower---with the
extreme case that no turns were possible against oncoming traffic of
speed zero.
}

\subsection{Signalized intersections}
In TRANSIMS, we distinguish between signalized intersections and
unsignalized intersections because they are modeled differently in
TRANSIMS.  In signalized intersections, the priorities are changing in time
and regulated by signals.  In unsignalized intersections, the priorities
are fixed.

When a simulated vehicle approaches a {\em signalized\/} intersection, the
algorithm first decides if, according to its current speed, it potentially
wants to leave the link, i.e.\ its current speed (in cells per update) is
larger than or equal to the remaining number of cells on the
link.\footnote{%
Vehicles may accelerate or slow down before they actually reach the
intersection.  See below.
} If a vehicle wants to leave the link, the algorithm checks the
``traffic control'', which determines if the vehicle can leave the
link.  If it encounters a red light, it can {\em not\/} leave the link
and no further action is taken.  If it encounters a protected (green
arrow) or caution (yellow) signal, the vehicle is allowed to enter the
intersection.  If it encounters a permitted signal (green, for example
permitted left turn against oncoming traffic), the vehicle checks all
interfering lanes for the gap that is larger or equal to 3~times the
oncoming vehicle's velocity (see Subsec.~\ref{unprotected} above).

If the movement into the intersection is accepted, the vehicle is moved
into an ``intersection queue''; there is one queue for each incoming lane.
This queue models vehicle behavior inside an intersection.  The vehicle
gets a ``time stamp'', before which it is not allowed to leave the
intersection; this time stamp is representative for the duration of the
movement through the intersection.  The intersection queues have finite
capacity; once they are full, no more vehicles are accepted and the
vehicles start to queue up on the link.  This models
the finite vehicle storing capacity of an intersection.

Once a vehicle is ready to leave the intersection, it moves to the first
cell on the destination link if available.\footnote{%
Algorithmically, it only ``reserves'' a cell.  See below.
} The speed of the vehicle is not changed when it is in the intersection
queue so it exits on the destination link in the first cell with the same
velocity that it had when it entered the queue.  

Note that vehicles turning against interfering traffic make their
decision to accept the turn when they {\em enter\/} the intersection queue,
not when they leave it.  This can have the effect that a vehicle enters the
intersection queue when there is no oncoming traffic, but, because of other
vehicles ahead of it in the same queue, cannot make its turn immediately.
Yet, since the turn was already accepted, it will be executed as soon as
all vehicles ahead in the same queue have cleared the queue and a cell on
the destination link is available. The turn can occur during oncoming 
traffic. So in some sense vehicles will go ``through'' each other.
Yet, note that on average the result is still correct. The approach
described above will not let more vehicles through the intersection than a
gap acceptance calculated when {\em leaving\/} the intersection queue.  The
above logic was chosen for simplification purposes since unsignalized
intersections (see below) do not have queues and thus {\em need\/} to make
their acceptance decisions when entering the intersection.

\subsection{Unsignalized intersections}
Unsignalized intersections in TRANSIMS have no internal queues, i.e.\
vehicles go right through them.\footnote{%
Again, technically the vehicles only reserve cells on the destination
links.  The actual move through the intersection happens later and can
also be postponed if after the velocity update the vehicle actually
does {\em not\/} make it to the intersection.
} Also, vehicles leaving an unsignalized intersection go down the
destination link as far as prescribed by their velocity, not just into
the first cell as in the signalized intersections.  Apart from these
two differences, unsignalized intersections are similar to signalized
ones.

When a simulated vehicle approaches an unsignalized intersection, the
algorithm first decides if, according to its current speed, it potentially
wants to leave the link, i.e.\ its current speed (in cells per update) is
larger than or equal to the remaining number of sites on the link.  If a
vehicle wants to leave the link, the algorithm checks the ``traffic
control'', which determines if the vehicle can leave the link.  Currently
occuring traffic controls are: no control, yield, and stop.  

If a ``no control'' is encountered, the vehicle is moved to its destination
cell without any further checks.  For example, if a vehicle has a velocity
of 5 cells per update and 2 more cells to go on its link, then it attempts
to go 3 cells into the destination link.  If that cell is already reserved
(either by another ``reservation'' or by a real vehicle), then the next
closer cell is attempted, etc., until the algorithm either finds an empty
cell or returns that the destination lane is full.  ``No control'' is
usually used for the major directions, i.e.\ for the lanes which have
priority.

If a ``yield'' is encountered, the vehicle checks the gap on all
interfering lanes.  According to the same rules as above, on all interfering
lanes the gap needs to be larger or equal three times the first
vehicle's speed on that lane.  If the movement is accepted, the destination
cell is selected according to the same rules as with the ``no control''
case.

If it encounters a ``stop'', the vehicle is brought to a stop.  Only
when the vehicle has a velocity of zero for at least one time step on
the last cell of the link, is it allowed to continue.  If the result
of the regular velocity update indeed accelerates the
vehicle,\footnote{%
I.e.\ there is a probability of $1-p_{noise}$ that the vehicle will
not accelerate in the given time step.
} then it
attempts to go through the intersection.  On all interfering lanes the
gap, according to the same rules as above, needs to be larger or equal
to three times the first vehicle's speed on that lane.  If the
movement is accepted, a vehicle coming from a stop sign will always go
to the first cell on the destination link (if empty) and will have a
velocity of one.

\subsection{Parking locations}
In the current TRANSIMS microsimulation, vehicular trips start and end at
parking locations.  Each link in the microsimulation, except for freeway
ramps, freeway links, and some ``virtual'' links such as centroid
connectors, has at least one parking location.  Parking locations thus
represent the aggregated parking options on that link.  Parking locations
have rules about how vehicles enter and exit the simulation:\begin{itemize}

\item
Each vehicle in TRANSIMS has a complete route plan, together with a
starting time.  At the starting time, the vehicle is added to a queue
of vehicles that want to leave the same parking location.  When the
vehicle is the first one in the queue, it attempts to enter the link.
The acceptance logic is in spirit similar to the logic of the
unsignalized intersections, i.e.\ vehicles check the available gap and
make their decision based on that.  Parking accessory logic is not the
focus of the current paper, and since that logic may change in
TRANSIMS in the near future and we also expect no influence on the
results presented here, we omit further technical details.

\item
A vehicle that has reached its destination parking location according to
its plan will leave the microsimulation.  It is simply removed from the
traffic.

\end{itemize}

\subsection{Parallel logic}
TRANSIMS is designed to run on parallel computers, such as coupled
workstations, desktop multi-processors, or supercomputers.  The
parallelization approach used for the microsimulation is a geographical
distribution, i.e.\ different geographical parts of the simulated area are
computed on different CPUs.  

The current TRANSIMS microsimulation has these boundaries always in the
middle of links.  This is done in order to keep the complexity of the
parallel computing logic as far away as possible from the complexity of the
intersection logic.

Information needs to be exchanged at the boundaries several times per
update in order to keep the dynamics consistent.  For example, if a vehicle
changes lanes and end up close in front of another one, that other one is
probably forced to brake.  Now, if the lane changing vehicle is on one CPU
and the following one on another, one needs to communicate the lane
change.  This will be called ``Update boundaries'' in the following section.

\subsection{Complete scheduling}
For a complete transportation microsimulation, we need to specify when
movements are accepted, and also how conflicts are resolved.  For
example, vehicles simultaneously attempting to change lanes into the
middle lane represent such a conflict.  Another conflict is two
vehicles from two different links competing for the same site on the
destination link.

The complete update of the current TRANSIMS microsimulation is as follows.
Assume that the state at time $t$ is the result of the last update.  Let
$t1$, $t2$, etc.\ be intermediate partial time steps.\begin{enumerate}

\item
Vehicles which are ready to leave intersection queues from signalized
intersections reserve cells on outgoing lanes.  They only attempt to
reserve the first cell on the link; their velocity is the same as it
was when they entered the intersection.  When the cell is occupied
(either by another ``reservation'' or by a vehicle), then the vehicle
cannot leave the intersection.  Note that there can be a conflict
between different queues for the same destination cell.  The current
solution in TRANSIMS is that queues are served on a first come first
served basis in some arbitrarily defined way, i.e.\ a queue which
happens to be treated earlier in the microsimulation has a slightly
higher chance of unloading its vehicles. --- Result: $t_1$
information.

\item 
Vehicles change Lanes.  Use information from time $t_1$ to calculate
situation at time $t_2$.

\item
Exit from Parking.  Results in $t_3$ information.

\item
Exchange boundary information for parallel computing.

\item
Non-signalized intersections reserve sites on target lanes.  Note that there
can be a conflict of two incoming links competing for the same destination
cell.  The current solution in TRANSIMS is that links are served on a first
come first served basis, i.e.\ a link which happens to be treated earlier
in the microsimulation has a slightly higher chance of unloading its
vehicles.  Note that this conflict only happens between minor links.  Major
links never compete for the same outgoing link except when there is a
network coding error; and for the competition between major and minor
links, the major link always wins because of the interfering lanes
conditions.\footnote{%
Note that the situation slightly different when the speed of the
vehicle on the major link is zero---see below.
} Result: $t_4$ information.

\item
Calculate speeds and do movements.  If a vehicle scheduled for an
intersection does not go through the intersection as a result of the
velocity update, the reservation is cancelled. Vehicles which go
through unsignalized intersections have $p$ set to zero, i.e.\ if it
turns out that the result of the velocity update indeed brings them
into the intersection, they need to go to the site on the destination
lane which was reserved earlier.  Result: $t_5 = t+1$ information.

\item
Exchange boundary information and migrate vehicles for parallel computing.

\end{enumerate}

\section{Standardized flow test suite for simulation models}
In order to control the effect of driving rules, TRANSIMS provides
controlled tests for traffic flow behavior.  These tests are
simplified situations where elements of the microsimulation can be
tested in isolation.  This test suite uses the standard
microsimulation code in the same way it is used for full-scale
regional simulations, and it also uses the same input and output
facilities: The test network is currently defined via a table in an
oracle data base, in the same format as the Dallas/Fort Worth network
is kept.  Input of vehicles is, following individual vehicle's plans,
via parking locations, the same way vehicles enter regional
simulations.  Output is collected on certain parts of the network on a
second-by-second basis, the same way it can be collected for regional
microsimulations.  The collected output is then post-processed to
obtain the aggregated results presented in this paper.

The test cases we look at in this paper are the following (see also
Fig.~\ref{defs}~(e)):\begin{itemize}

\item
One-lane traffic, in order to see if car following behavior generates
reasonable fundamental diagrams.

\item
Three-lane traffic, in order to see if the addition of passing lane changing
behavior still generates reasonable fundamental diagrams, and in order
to look at lane usage.

\item
Stop sign, yield sign, and left turns against oncoming traffic, in
order to see it the logic for non-signalized intersections generates
acceptable flow rates.

\item
A signalized intersection, in order to see of we obtain reasonable
flow rates, and in order to check lane changing behavior for plan
following purposes.

\end{itemize}

\subsection{Measured quantities}
We look at three minute averages of the following
quantities:\begin{itemize}

\item
{\bf Local Flow.}  Flow~$q$ is defined as usual by:
\begin{eqnarray}
{\it q\/} = \frac{\it N\/}{\it T} \;\;\;\;\;\; [vehicles/hour]  \nonumber
\end{eqnarray}
{\it N} is the number of cars which pass a certain site at a time period
{\it T\/}.

\item
{\bf Local Density.}  Density is in principle easily defined, $\rho =
N / L$, where $N$ is the number of vehicles on a piece of roadway of
length~$L$.  Yet, given current sensoring technology, this is not easy
to achieve since one would need a sensor which counts, say once a
second, cars on a predefined stretch of length~$L$ of the roadway.
For that reason, empirical papers sometimes resort to occupancy, which
is the fraction of time a given sensor has been occupied by a vehicle.
Current TRANSIMS measures density according to its original
definition, i.e., once a time step, we count the number of vehicles on
a stretch of roadway of $L=5$~sites $=5 \times 7.5$~m $=
37.5$~m.\footnote{%
The ``magical'' number of $L=5$~sites is equal to the maximum velocity
of $v_{max} = 5$~sites/update.  This ensures that each vehicle is
counted at least once.
} We add these counts for $k=180$ measurement events and then divide
the resulting number by $L$ and by $k$:
\begin{eqnarray}
{\it \rho\/} = \frac{\it N\/}{\it k * L}  \nonumber
\end{eqnarray}
The result can be scaled to convenient units, for example ``vehicles
per km''.

Note that this way of computing density averages the counts over a
length of 37.5~m, which is longer than the usual sensor extensions.
The effect of this should be systematically studied.

\item
{\bf Local velocity.}  It is well known that one can measure velocity
either analogous to our flow definition (local velocity) or analogous
to our density definition (space-averaged velocity).  Under
non-stationary conditions, the measurements give different results,
since, for example, the first definition never counts vehicles with
velocity zero.  Local velocity is easier to measure in practice; the
space-averaged velocity is easier to interpret since it is equal to
the travel velocity and it is also the velocity which needs to be used
in the fundamental relationship between flow, density, and velocity,
$q = \rho \cdot v$.  Since in a simulation model, both are similarly
easy to measure, we measure the more useful travel velocity.  Once a
time step, we sum up the individual velocities of all vehicles on a
stretch of roadway of $L=5$~sites $=5 \times 7.5$~m $= 37.5$~m.  We
add these sums for $k=180$ measurement events and then divide the
resulting number by $N$ and by $k$, where $N$ is the same number as
obtained during the density measurement above:
\[
v = {\sum v \over k * N}
\]

\item
{\bf Lane usage.}  Lane usage of a particular lane is the number of
cars on this lane divided by the number of cars on all lanes.  It can
be computed as:
\[
f_i = {\rho_i \over \sum_{j=1}^n \rho_j \cdot n} \ ,
\]
where $i$ is the lane we look at and $n$ is the number of lanes.

\end{itemize}

\subsection{Test networks}
%
%
Essentially two test networks are used: a circle of 1\,000~sites $=$
0.75~km in various configurations, and a simple signalized
intersection.  Most of the test are run on the circle networks.  The
circle can have one or two or three lanes.  In all tests, the circle
is slowly loaded with traffic via a parking location at $x=1$~sites.
Velocity, flow, and density are measured on $486 \le x \le 490$,
thus generating the fundamental diagrams for one-lane, two-lane, and
three-lane traffic.  Since the circle gets slowly loaded, the
complete fundamental diagram is generated during one run.

For testing yield signs and stop signs, an incoming lane is added on
the right side of traffic at $x=501$ (i.e.\ the first cell for the
incoming traffic is 501).  The characteristics of the incoming
traffic is measured on a measurement box on the last 5~sites of the
incoming lane.  The incoming lane is operated at maximum flow, i.e.\
with as many vehicles as possible entering at its beginning.  The
incoming vehicles are removed at $x=900$ via a parking accessory.
The result of this measurement is typically a diagram showing the flow
of incoming vehicles on the y-axis versus the flow on the circle on
the x-axis.

For testing left turns against oncoming traffic, an opposing lane is
added so that it ends at $x=500$.  The traffic control here is again
a ``yield'' logic; the difference from before is that vehicles only {\em
traverse\/} the interfering traffic, they do not join it.

Last, a three-lane intersection approach is used.  The left lane makes
a left turn, the middle lane goes straight, the right lane makes a
right turn.  Incoming vehicles have plans about their intended
movement at the intersection and attempt to reach the corresponding
lane.  The intersection has signals with 1~minute green phase and
1~minute red phase. The typical output from this run is the flow of
vehicles which go through the intersection, and the number of vehicles
which cannot make their intended turn because they did not reach their
lane. 

The results are shown in Figs.~\ref{1lane}
to~\ref{unprotected-fig}.  Note that the HCM has the same
curves for stop sign and for yield sign, whereas we obtain higher
flows through yield signs, as should be the case.

\section{Some study results}
Most of the results presented here were generated with an experimental
code.  The disadvantage of an experimental code is that actual
implementation in the production version may still introduce changes
in the results due to small discrepancies.  The advantage is that
turnover (compile times, complexity of code, etc.) is much higher than
with a production version.  We used that advantage to test many
different rules.  In the following, we want to present a small
subsection of tests.

\begin{figure}
\centerline{\hfill
\epsfxsize0.5\hsize\epsfbox[47   197   553   604]{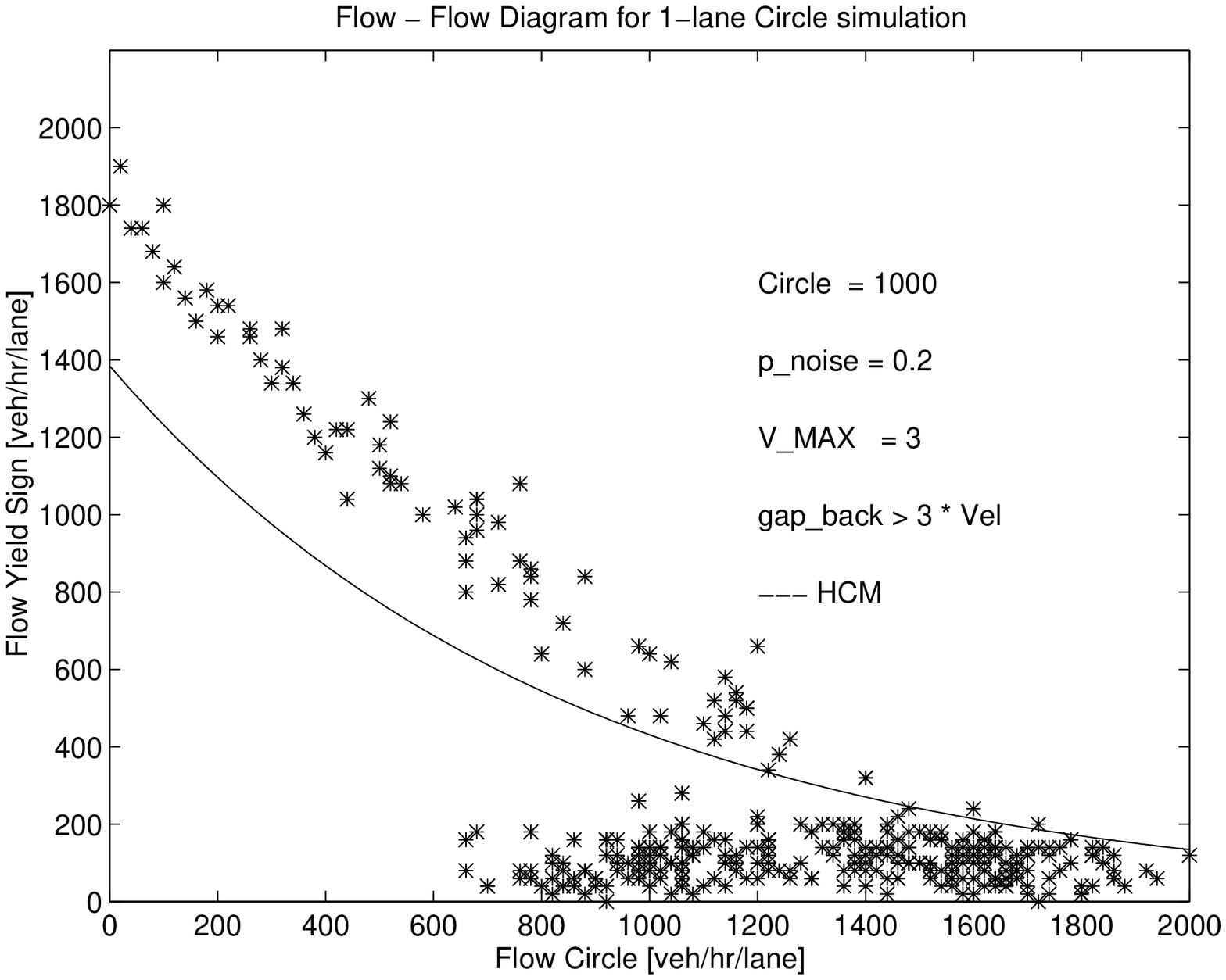}%
\epsfxsize0.5\hsize\epsfbox[47   197   553   604]{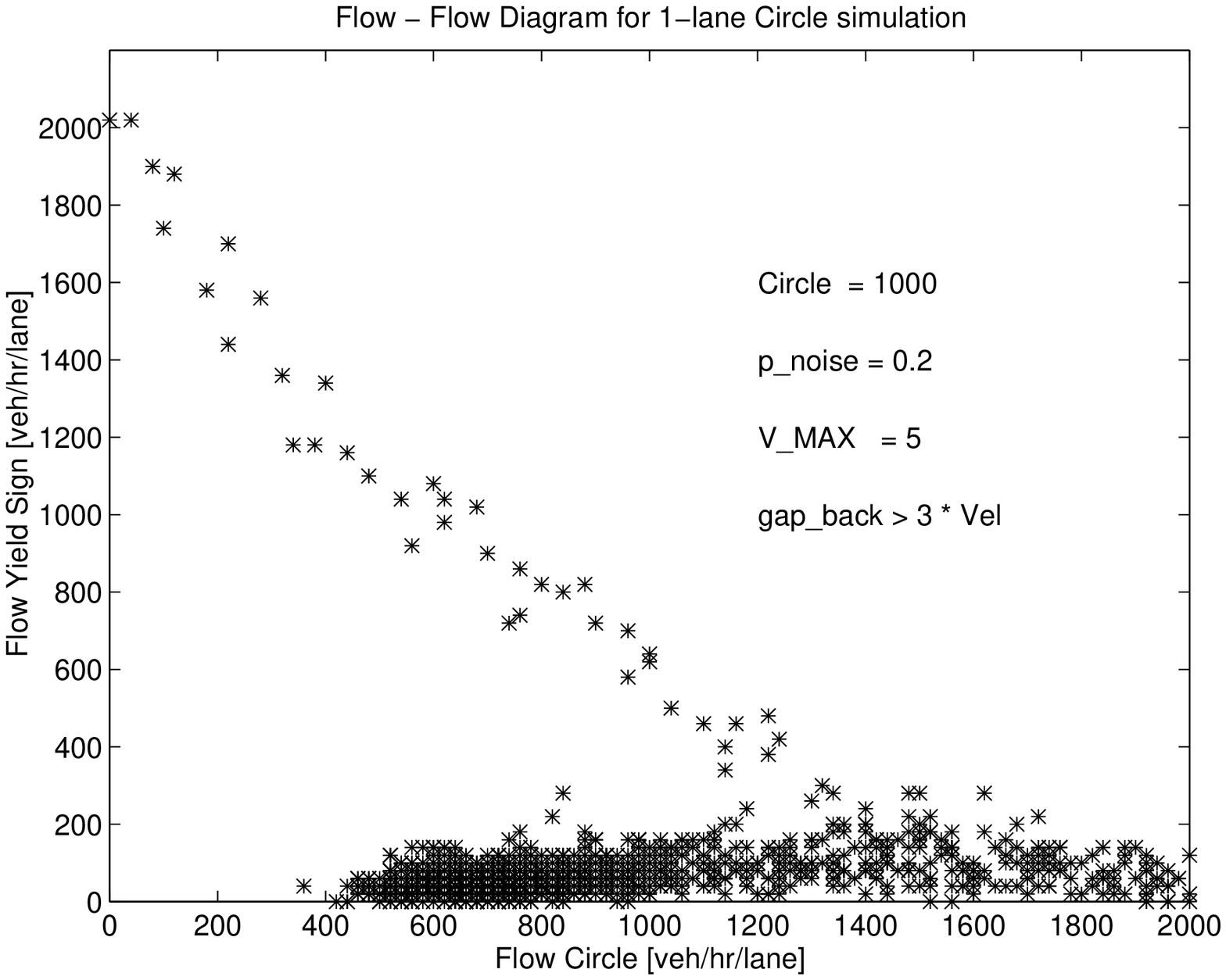}%
\hfill}
\centerline{\hfill
\epsfxsize0.5\hsize\epsfbox[47   197   553   604]{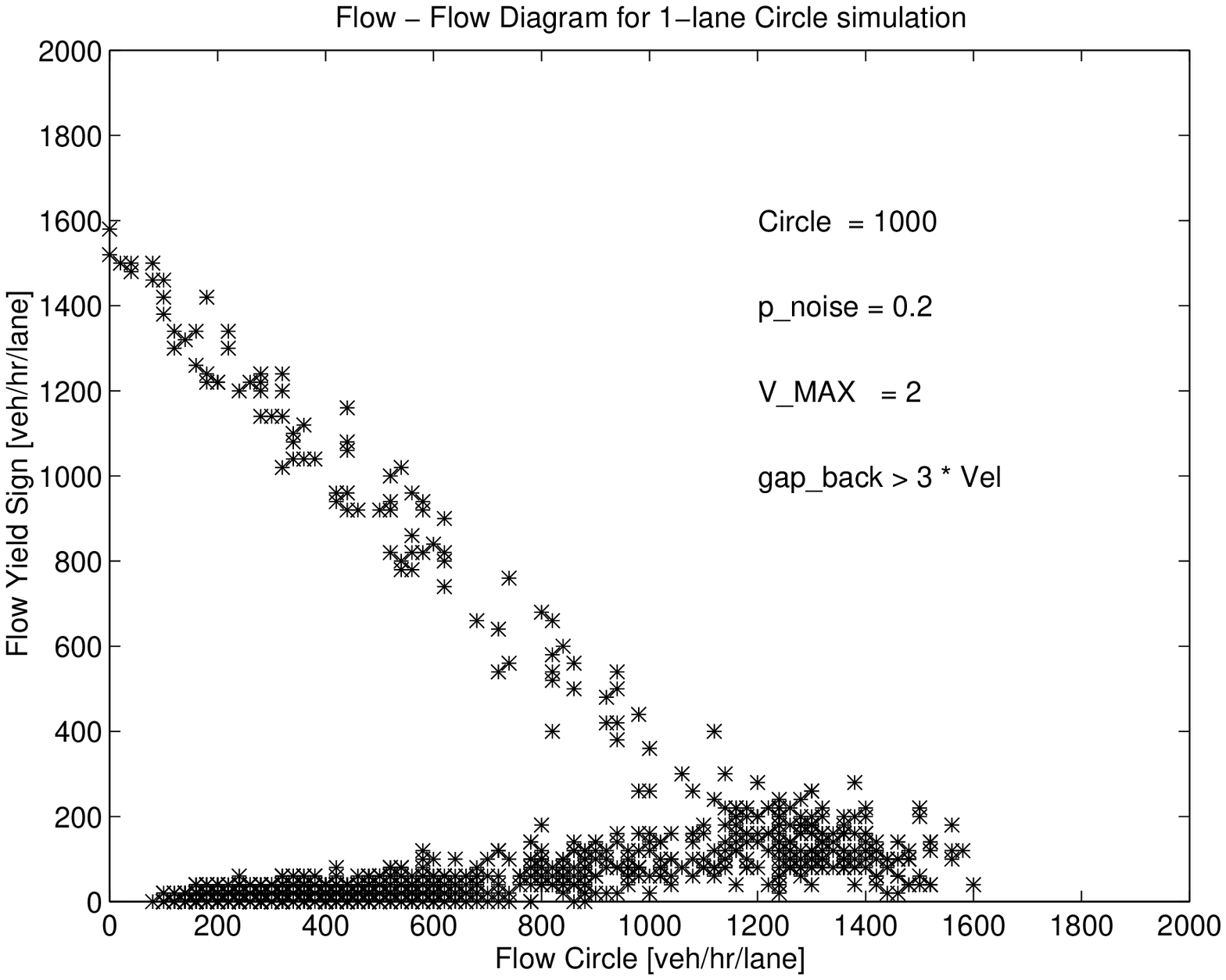}%
\epsfxsize0.5\hsize\epsfbox[47   197   553   604]{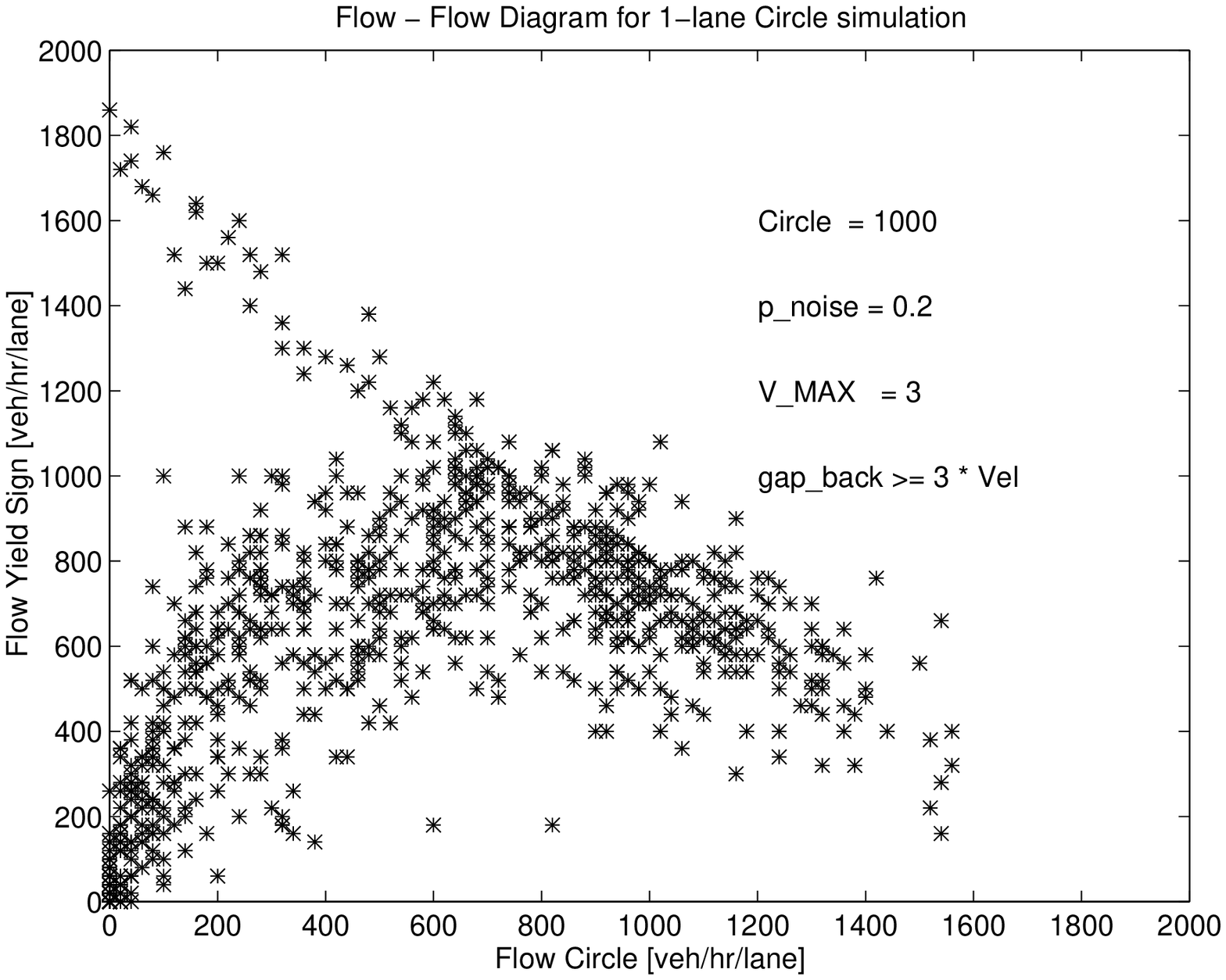}%
\hfill}
\centerline{\hfill
\epsfxsize0.5\hsize\epsfbox[47   197   553   604]{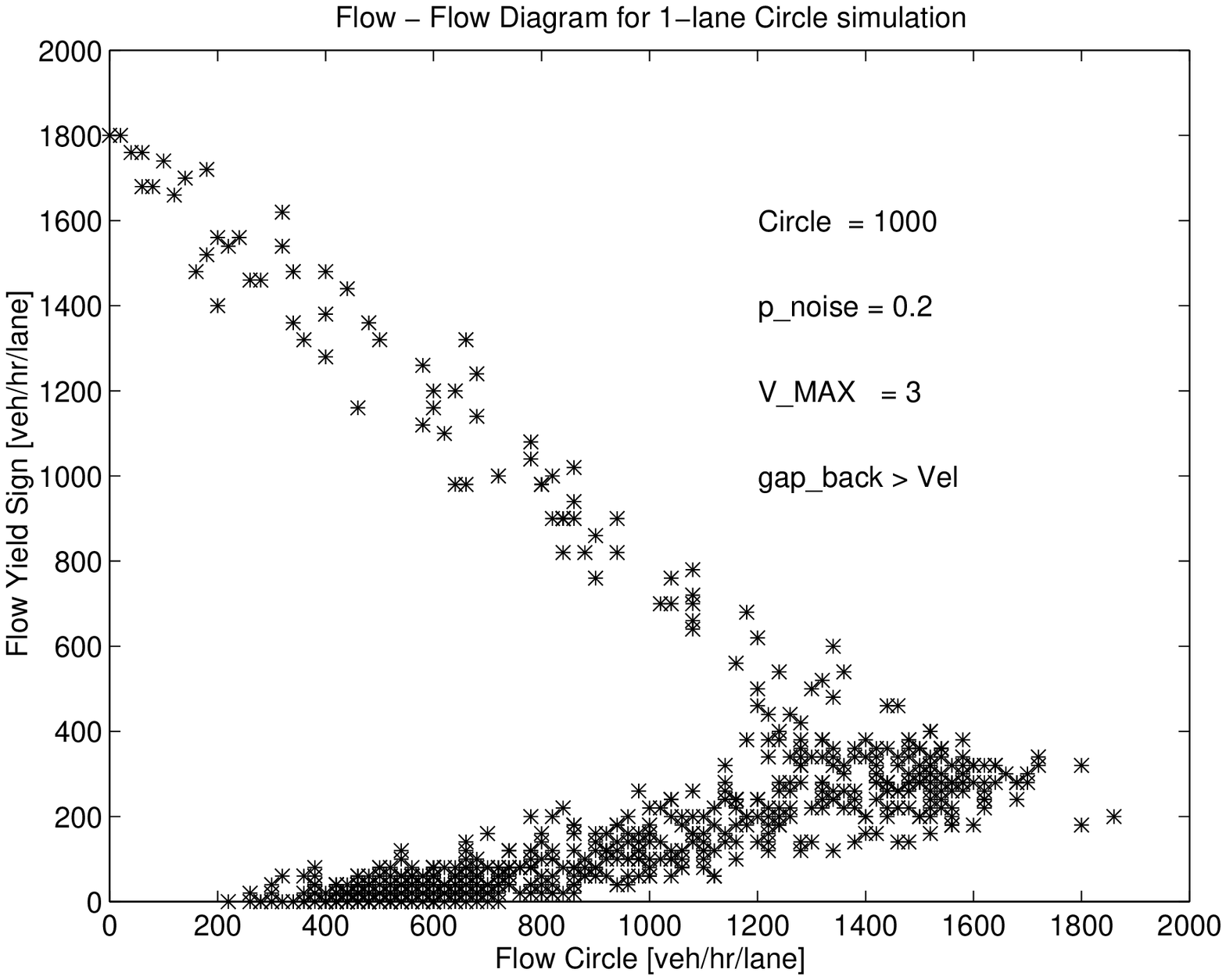}%
\epsfxsize0.5\hsize\epsfbox[47   197   553   604]{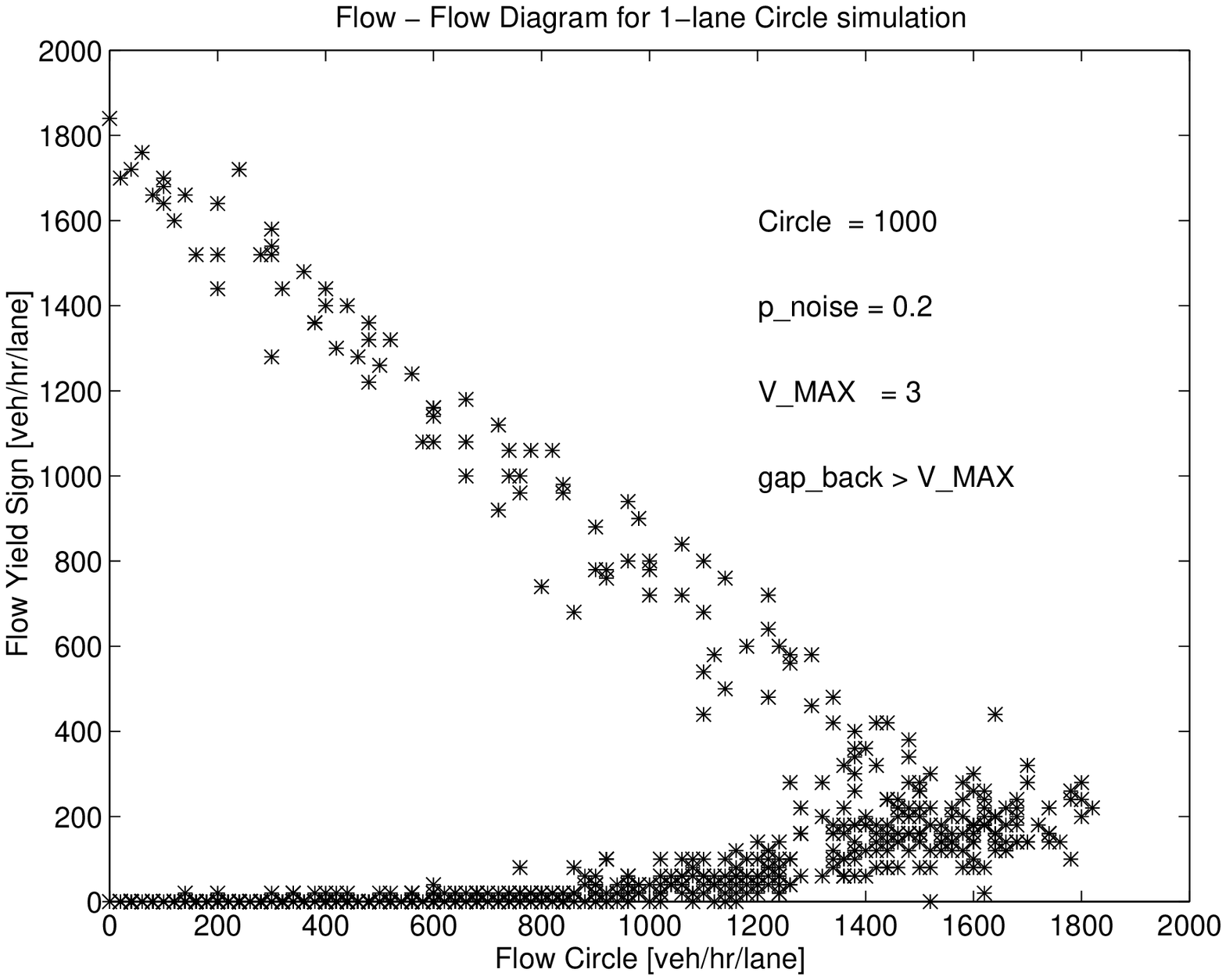}%
\hfill}
\caption{\label{study}%
Comparison between different rules for the case of a 1-lane minor road
controlled by a yield sign merging into a 1-lane major road. {\em
(a)\/}~Figure as shown earlier, i.e.\ ``accept if $gap > 3 v_{back}$''
and $v_{max}=3$.  {\em (b) -- (c)\/}~Effect of different maximum
velocities $v_{max}=5$ and $v_{max}=2$.  {\em (d)\/}~Effect of a
slightly different acceptance rule ``accept if $gap \ge 3 v_{back}$''
($v_{max}=3$).  {\em (e) -- (f)\/}~Effect of weaker gap acceptances
``accept if $gap > v_{back}$'' and ``accept if $gap > v_{max}$''
($v_{max}=3$).
}
\end{figure}

All results presented in this section refer to the situation of a
1-lane minor street merging into a 1-lane major street, with the
intersection control being a yield sign.  Fig.~\ref{study}~(a) repeats
the result from above for convenience.  Figs.~\ref{study}~(b)--(c)
show the result of different average free speeds (e.g. result of speed
limits) in the simulation (same speed limit for both streets).  A high
average free speed of approx.\ 130~km/h ($\approx 80$~mph, generated
by $v_{max}=5$), maybe a freeway merge, generates a flow of approx.\
2000~veh/hour/lane in the incoming lane when there is no traffic on
the major road.  From there, maximum incoming flow decreases
continuously.  Lower average free speeds of approx.\ 75~km/h (50~mph)
and 50~km/h (30~mph) generate lower maximum incoming flows and are
generally closer to the Highway Capacity Manual curve.  Yet, it should
also be clear from these curves that the flow on the minor road as a
function of flow on the major road is also a function of the speed
limit and not only of the gap acceptance, which is constant in all
three simulations.

Another series of experiments shows the effect of different acceptance
logics.  Fig.~\ref{study}~(d) shows, when compared to
Fig.~\ref{study}~(a), the difference between ``accept when $gap \ge 3
v_{back}$'' vs.\ ``accept when $gap > 3 v_{back}$''.  This seems like
a negligible difference in the rules; yet, the results are quite
different in the congested regime.  Whereas in the first, quite many
vehicles are able to get into the congested major road, in the second,
only few of them make it.  The difference is easiest explained by
looking at a vehicle of speed zero on the major road just in front of
the merge point, with space for a vehicle downstream of the merge
point.  With the first rule, a vehicle at the yield sign will accept
the move and move in front of the vehicle on the major road, in the
second case, it will not.  Both scenarios seem to be plausible to us;
only systematic measurements can probably resolve which one is better
for a simulation model.

A last series of experiments shows the effect of different values for
the gap acceptance.  Figs.~\ref{study}~(e) and~(f) show ``accept when
$gap > v_{back}$ and $gap > v_{max}$.  Clearly, more vehicles are
accepted, leading to a higher flow of turning vehicles as a function
of the flow on the major road.  Note that the flow via the yield sign
is never higher than 1800 minus the flow on the major road.  This
reflects the fact that the major road cannot have a higher flow than
1800~veh/h/lane (free speed approx 50~mph); traffic through the yield
sign can thus at most fill the major road to capacity.  This explains
why the much weaker gap acceptances to not produce even more
difference in the regime where the major road is uncongested.  The
situation is clearly different for unprotected turns {\em across\/}
instead of {\em into\/} traffic, as can be seen for the left turns in
the next section.

\section{Comparison to Case Study Logic}
The gap acceptance logic presented here and used in the current
TRANSIMS microsimulation is different from the logic used in the
``Dallas/Fort Worth Case
Study''~\cite{Beckman:etc:case-study,Nagel:Barrett:feedback}.  The
logic during that case study was: ``Accept an unprotected movement if
in all interfering lanes the gap is larger than $v_{max}=5$.''  This
means that at low flow rates on the major road, more turns were
accepted, whereas at high flow rates on the major road, less turns
were accepted.

\begin{figure}
\centerline{\hfill
\epsfxsize0.5\hsize\epsfbox[47   197   553   604]{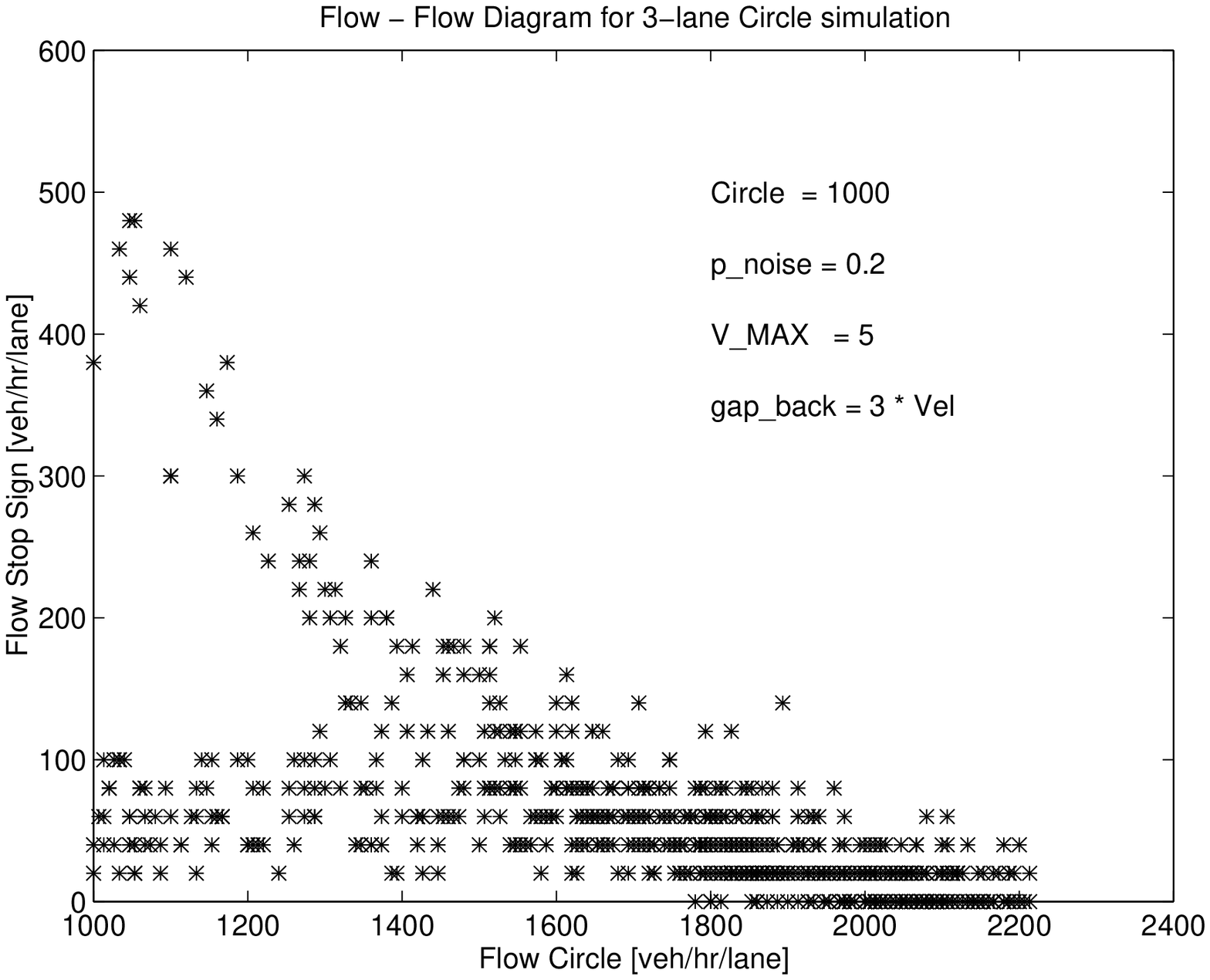}%
\epsfxsize0.5\hsize\epsfbox[47   197   553   604]{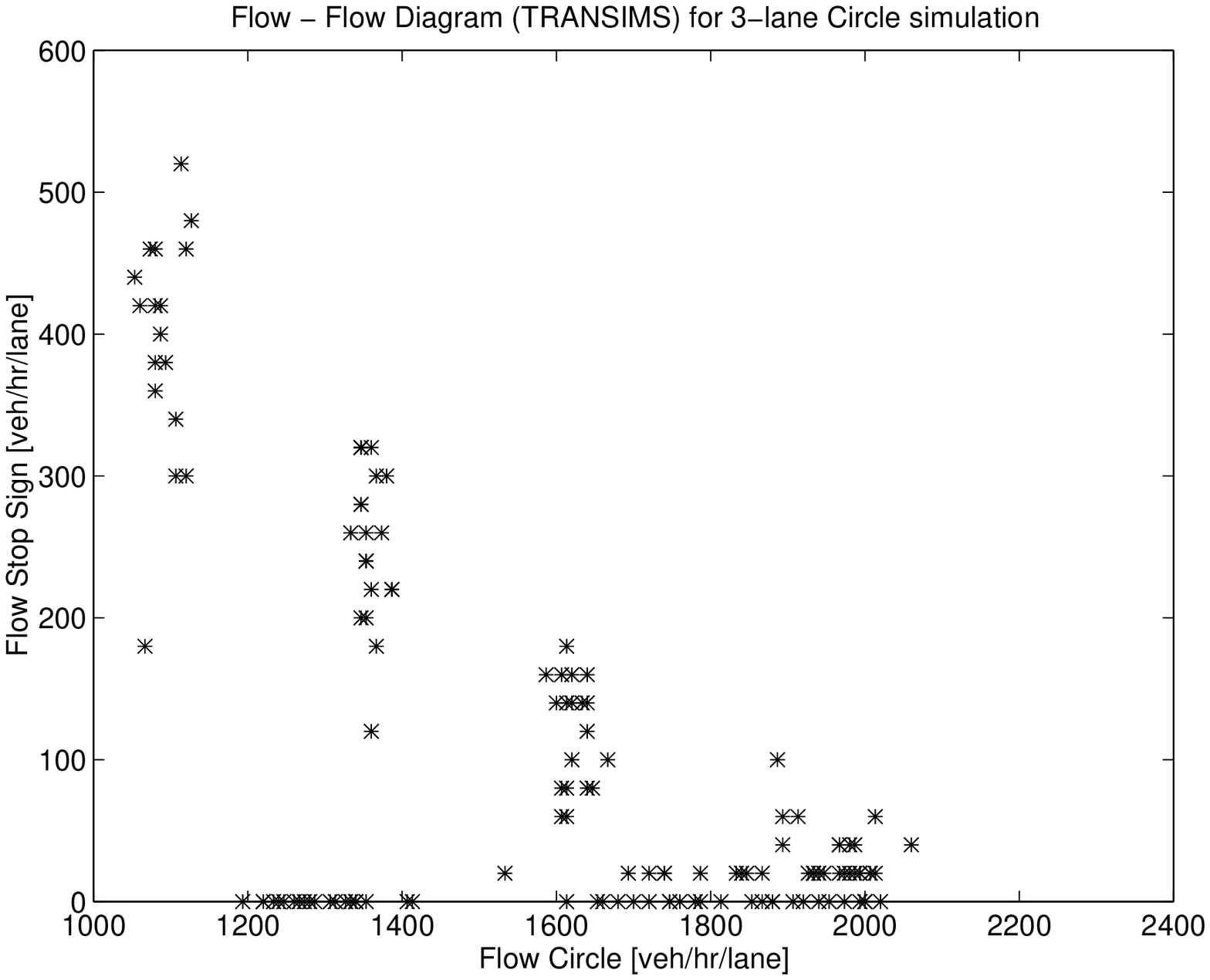}%
\hfill}
\centerline{\hfill
\epsfxsize0.5\hsize\epsfbox[47   197   553   604]{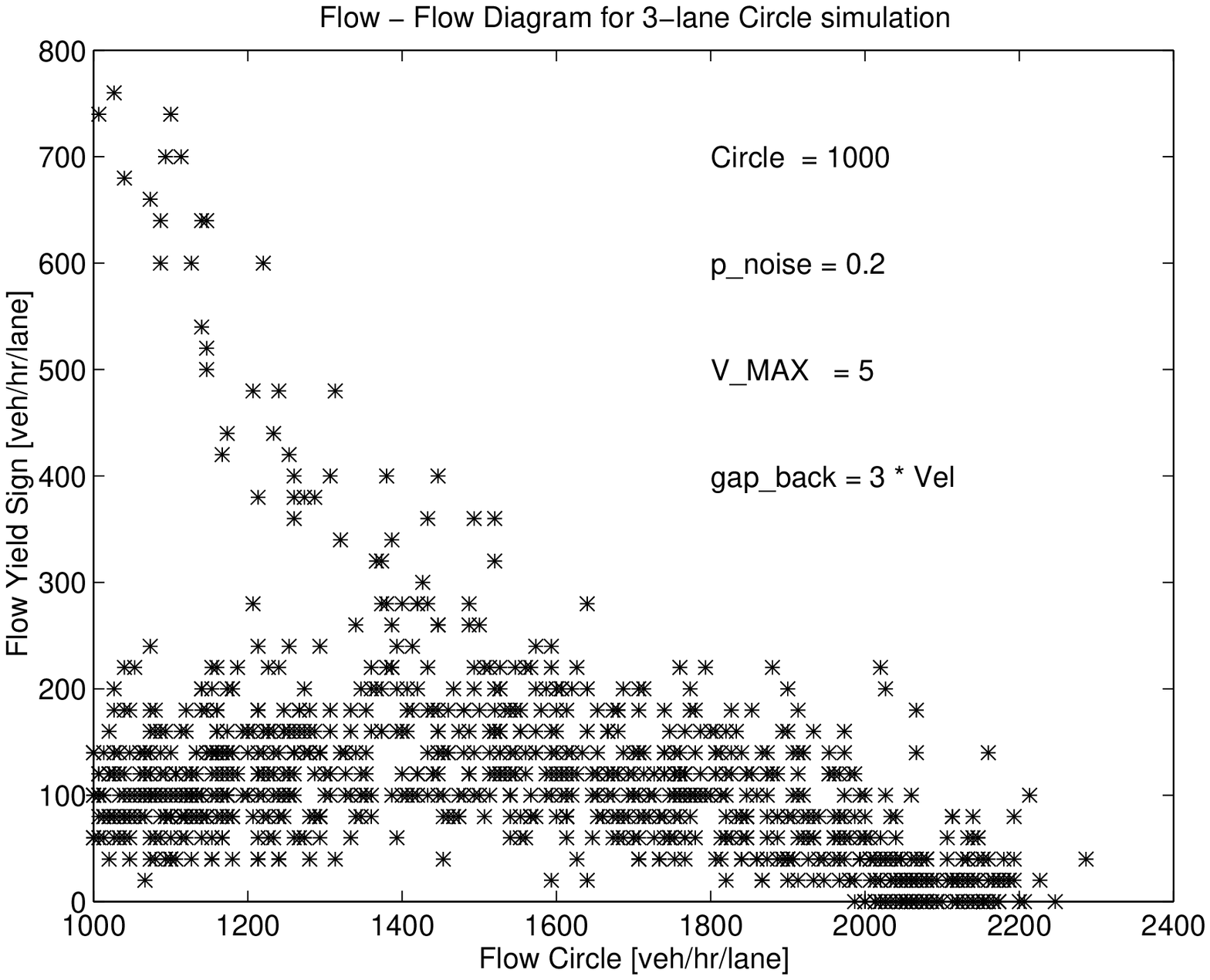}%
\epsfxsize0.5\hsize\epsfbox[47   197   553   604]{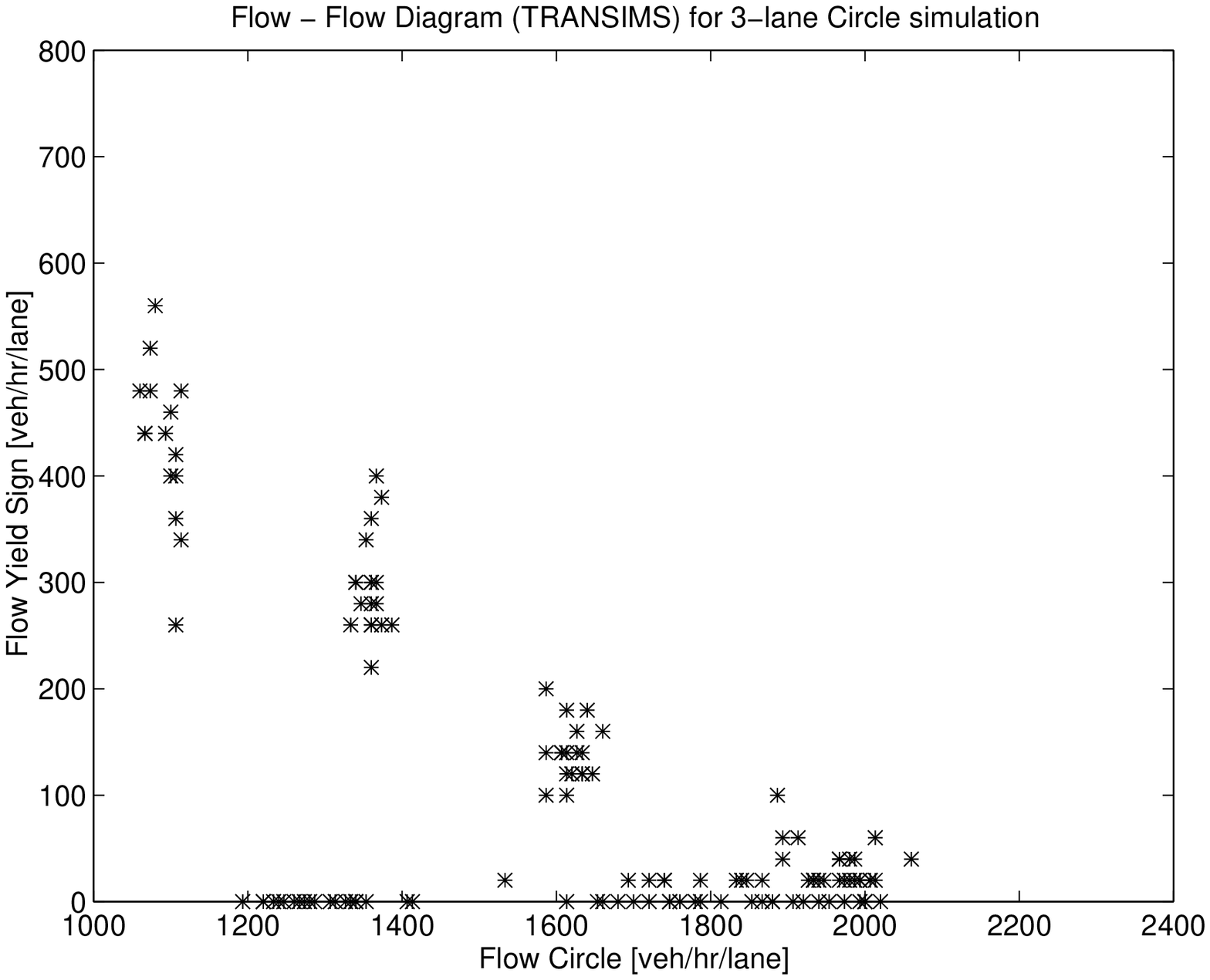}%
\hfill}
\centerline{\hfill
\epsfxsize0.5\hsize\epsfbox[47   197   553   604]{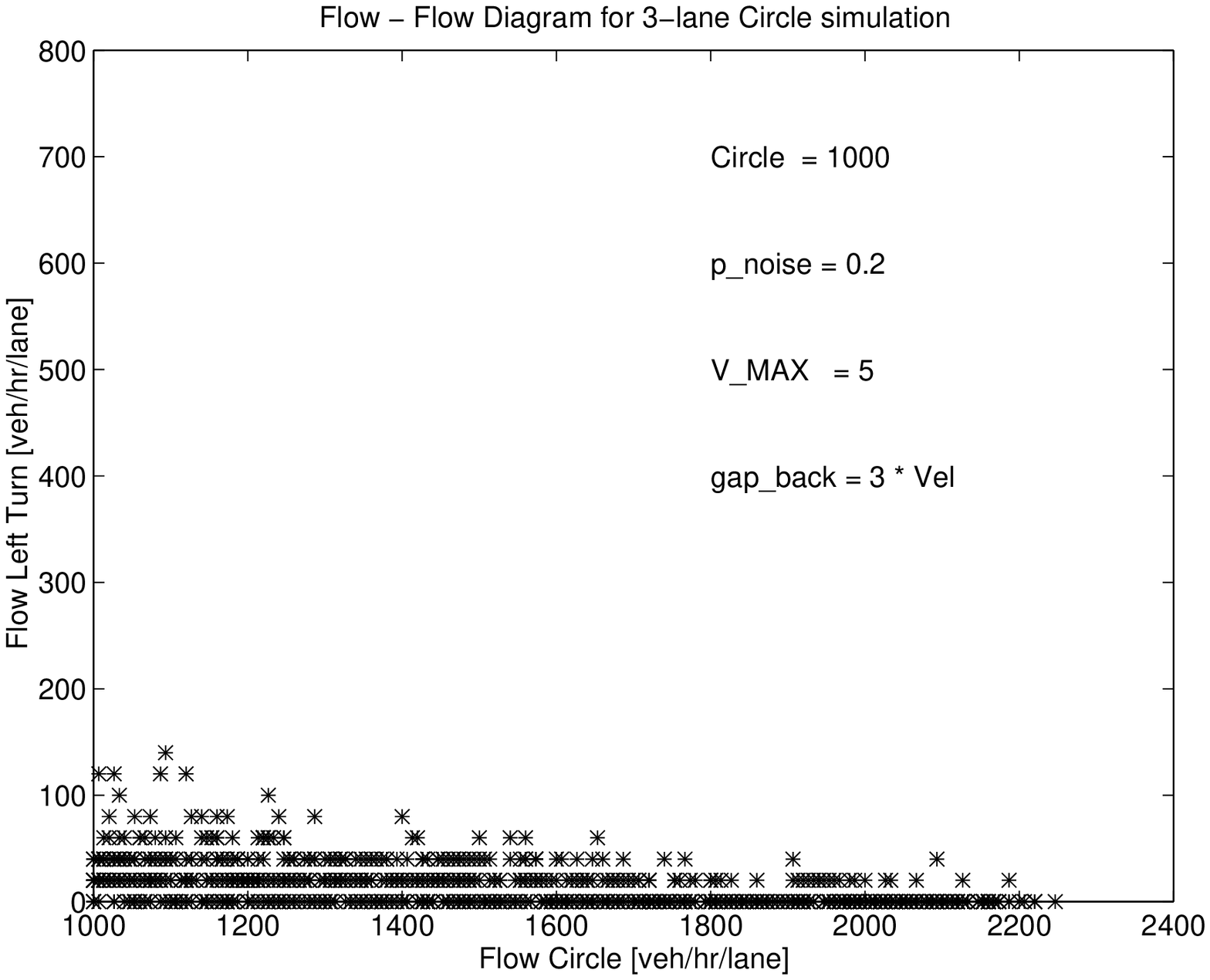}%
\epsfxsize0.5\hsize\epsfbox[47   197   553   604]{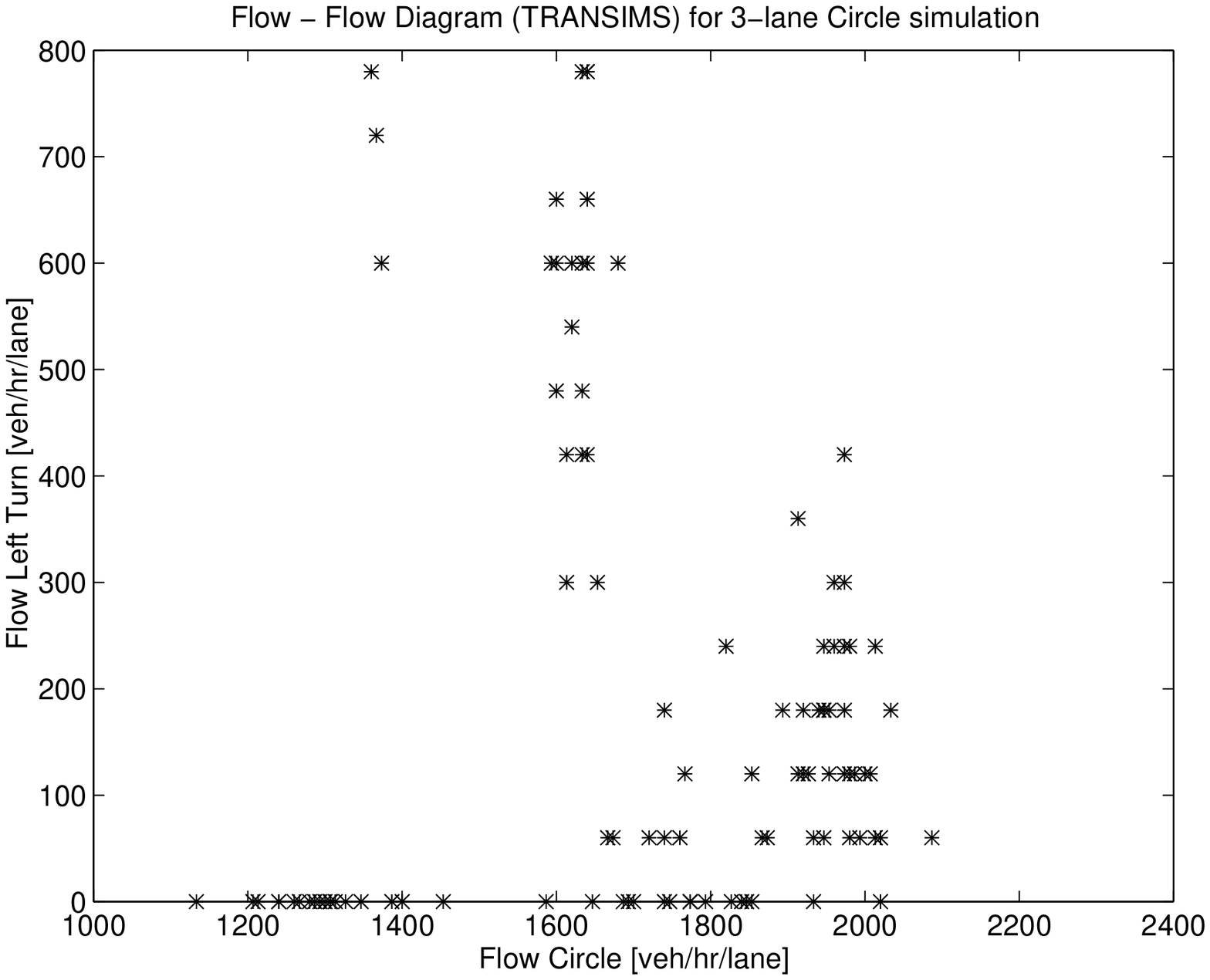}%
\hfill}
\caption{\label{comparison:3lane}%
Comparison between current TRANSIMS microsimulation gap acceptance
logic and the one used in the case study where the major road has
three lanes.  Flow through stop sign, yield sign, and unprotected left
turn into/across one-lane traffic on major road.  Left column: current
TRANSIMS microsimulation.  Right column: case study TRANSIMS
microsimulation.  Note that the results for the turns {\em into\/}
other traffic are not that much different whereas the result for the
turns {\em across\/} other traffic yields much higher flows with the
case study logic.
}
\end{figure}

Fig.~\ref{comparison:3lane} compares the results for the current
gap-acceptance logic and the one used in the case study for the case
where the major road is a 3-lane road.  Note that the results for the
turns {\em into\/} other traffic are not that much different
whereas the result for the turns {\em across\/} other traffic yields
dramatically higher flows with the case study logic.  This is due to
the fact that for turns {\em into\/} other traffic, there is a
capacity constraint of the form that the joint flows from the major
and the incoming road cannot exceed capacity of the major road.  Such
a constraint obviously does not exist for turns {\em across\/} the
major road.

\section{Summary and conclusion}
In transportation simulation models for larger scale questions such as
planning, the flow characteristics of the traffic dynamics are in some
sense more important than the microscopic driving dynamics of the
vehicles itself.  This becomes especially true since a ``complete''
representation of human driving is impossible anyway, both due to
knowledge constraints and due to computational constraints.  Yet,
calibrating a traffic simulation model against all types of desired
behavior (for example against all HCM curves and values mentioned in
this paper) seems a hopeless task given the high degrees of freedom.

TRANSIMS thus attempts to generate plausible macroscopic behavior from
{\em simplified\/} microscopic rules.  This paper described the more
important aspects of these rules as currently implemented or under
implementation in TRANSIMS.  Before we implement rules in the TRANSIMS
production version, we usually try to run systematic studies with more
experimental versions.  The results of the traffic flow behavior from
that study were presented.  Also, we showed the effects of some
changes in the rules for the example of a yield sign.  Finally, some
comparisons were made between the logic currently under implementation
and the logic used for the Dallas/Fort Worth case study.

\begin{figure}

\centerline{\hfill
\epsfxsize0.5\hsize\epsfbox[47   197   553   604]{ps_000.ps}%
\epsfxsize0.5\hsize\epsfbox[47   197   553   604]{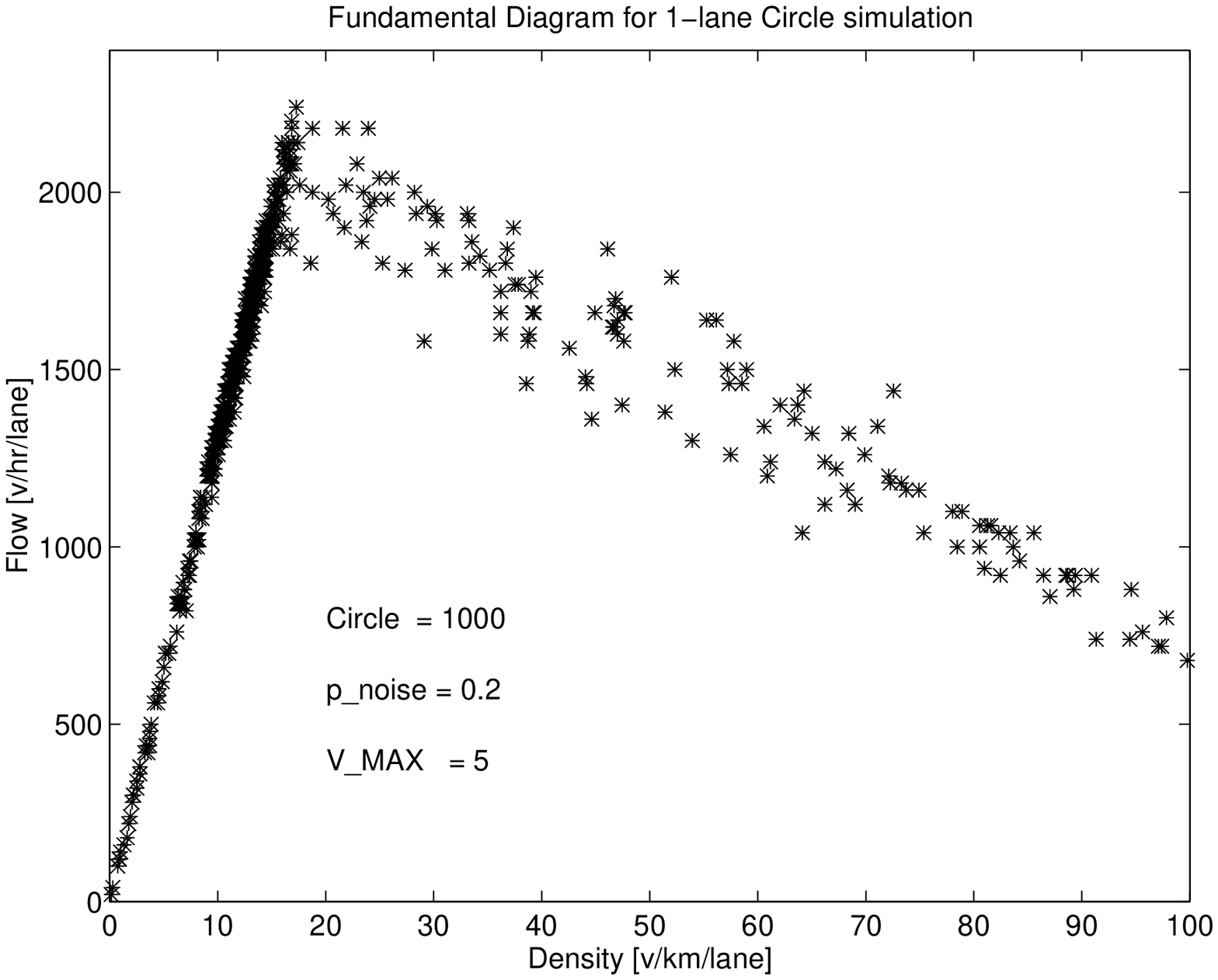}%
\hfill}
\centerline{\hfill
\epsfxsize0.5\hsize\epsfbox[47   197   553   604]{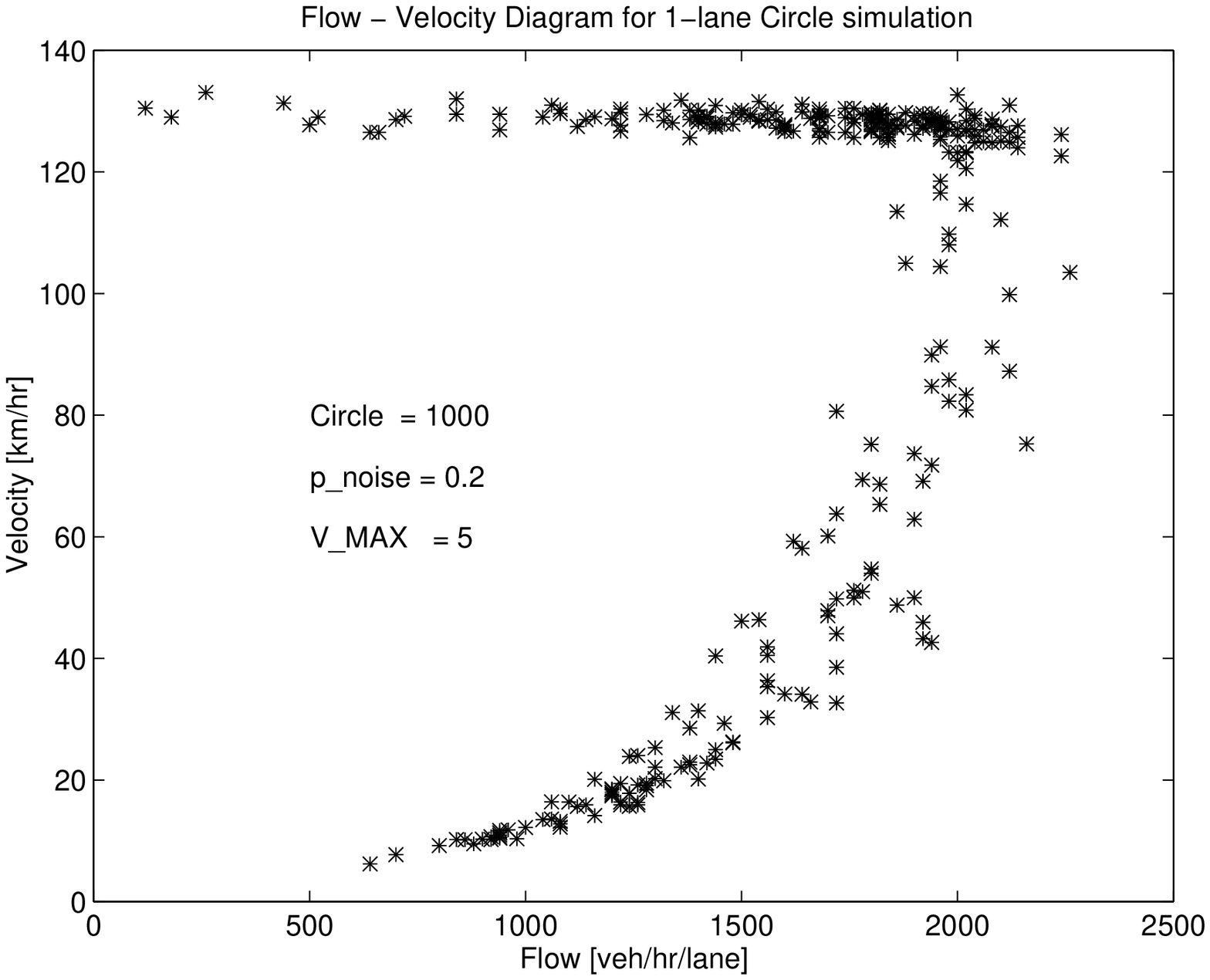}%
\epsfxsize0.5\hsize\epsfbox[47   197   553   604]{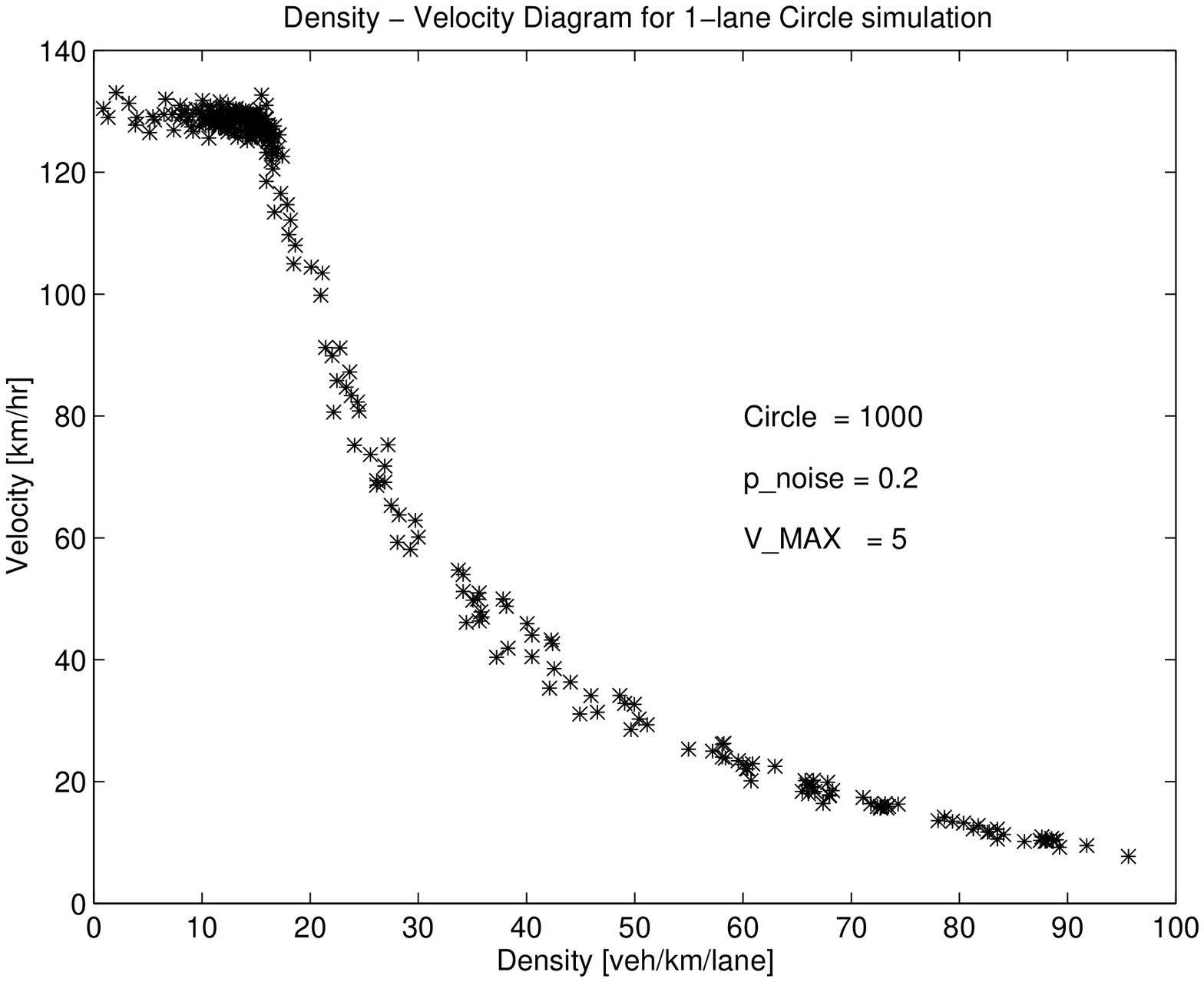}%
\hfill}
\centerline{(a)}

\vskip0.5cm

\centerline{\hfill
\epsfxsize0.5\hsize\epsfbox[47   197   553   604]{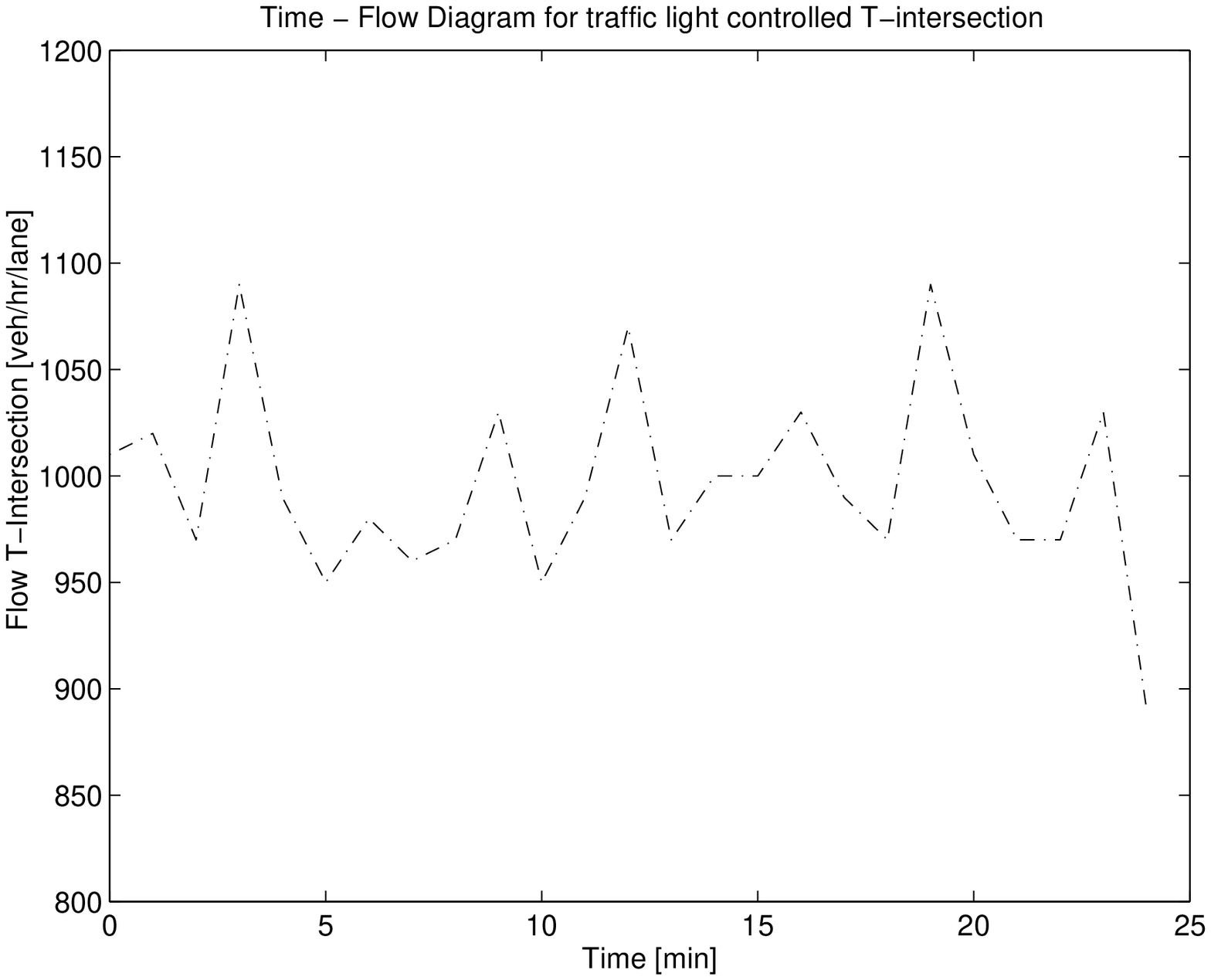}%
\hfill}

\centerline{(b)}

\caption{\label{1lane}
{\em (a)}~One-lane traffic: Flow vs.\ density, travel velocity vs.\ flow, and
travel velocity vs. density.
{\em (b)}~Number of vehicles going through the intersection and number
of vehicles ``off plan'' per green phase, re-scaled to hourly flow
rates per lane.
}
\end{figure}

\begin{figure}
\vspace*{\fill}
\centerline{\hfill
\epsfxsize0.5\hsize\epsfbox[47   197   553   604]{ps_000.ps}
\epsfxsize0.5\hsize\epsfbox[47   197   553   604]{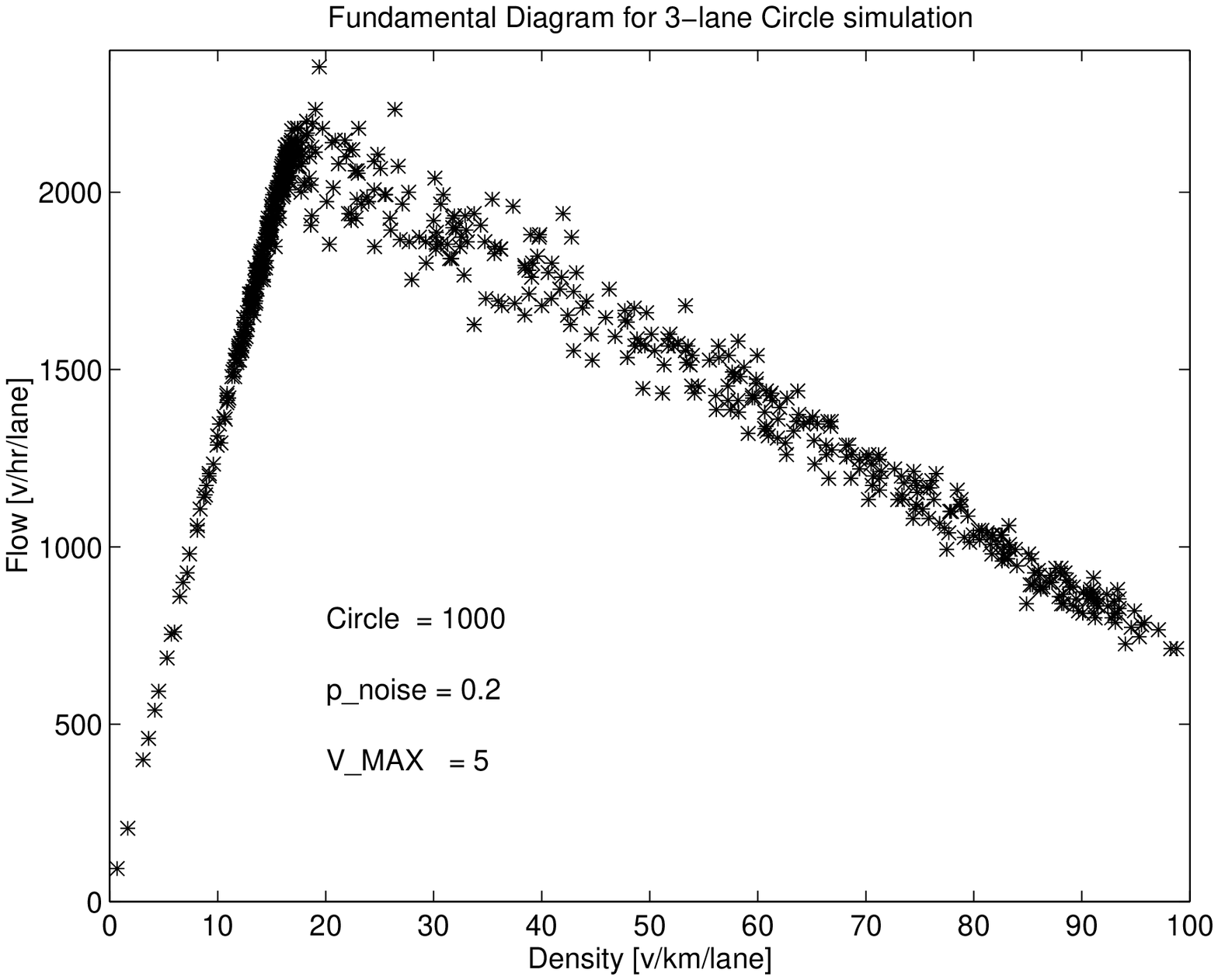}
\hfill}
\centerline{\hfill
\epsfxsize0.5\hsize\epsfbox[47   197   553   604]{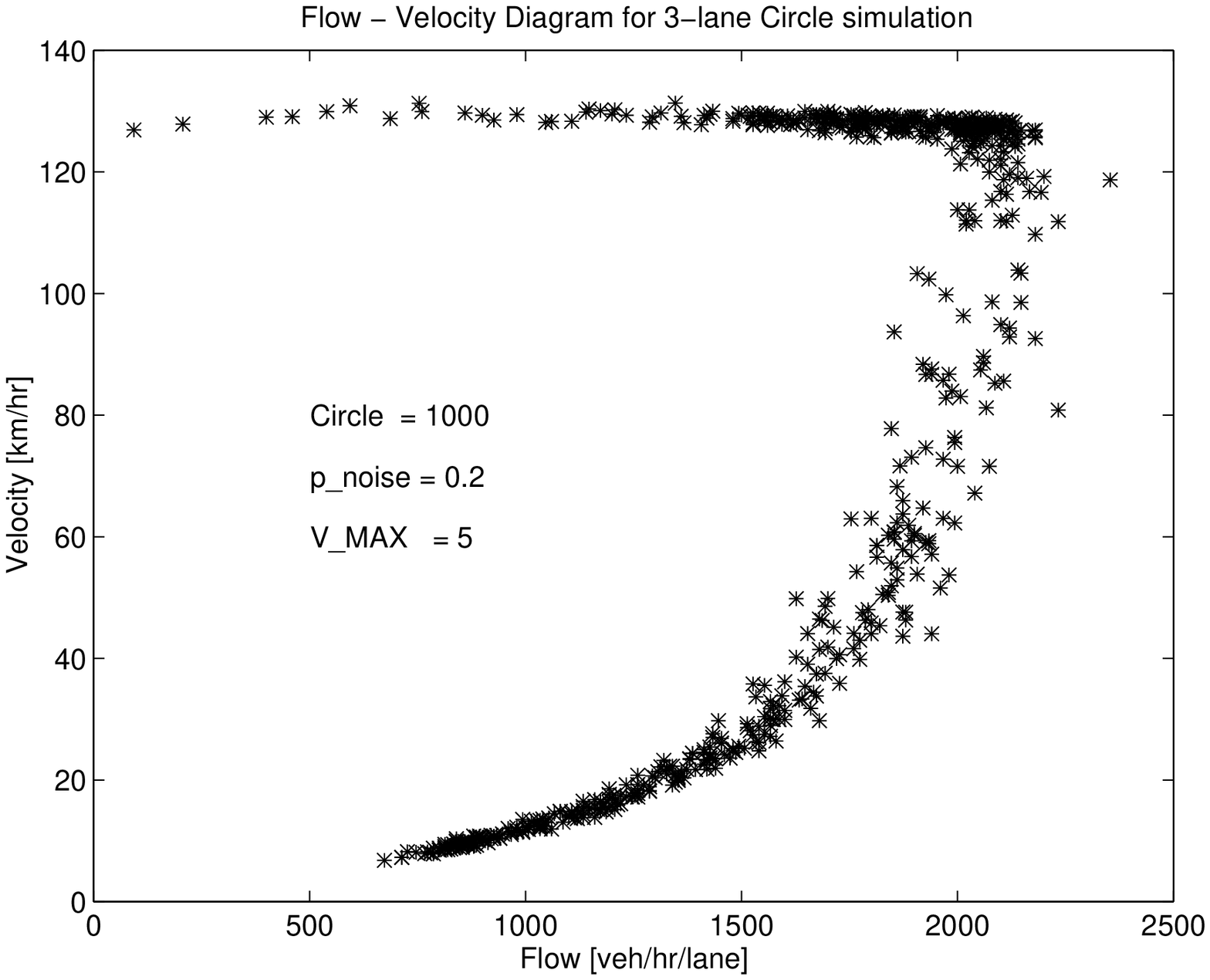}%
\epsfxsize0.5\hsize\epsfbox[47   197   553   604]{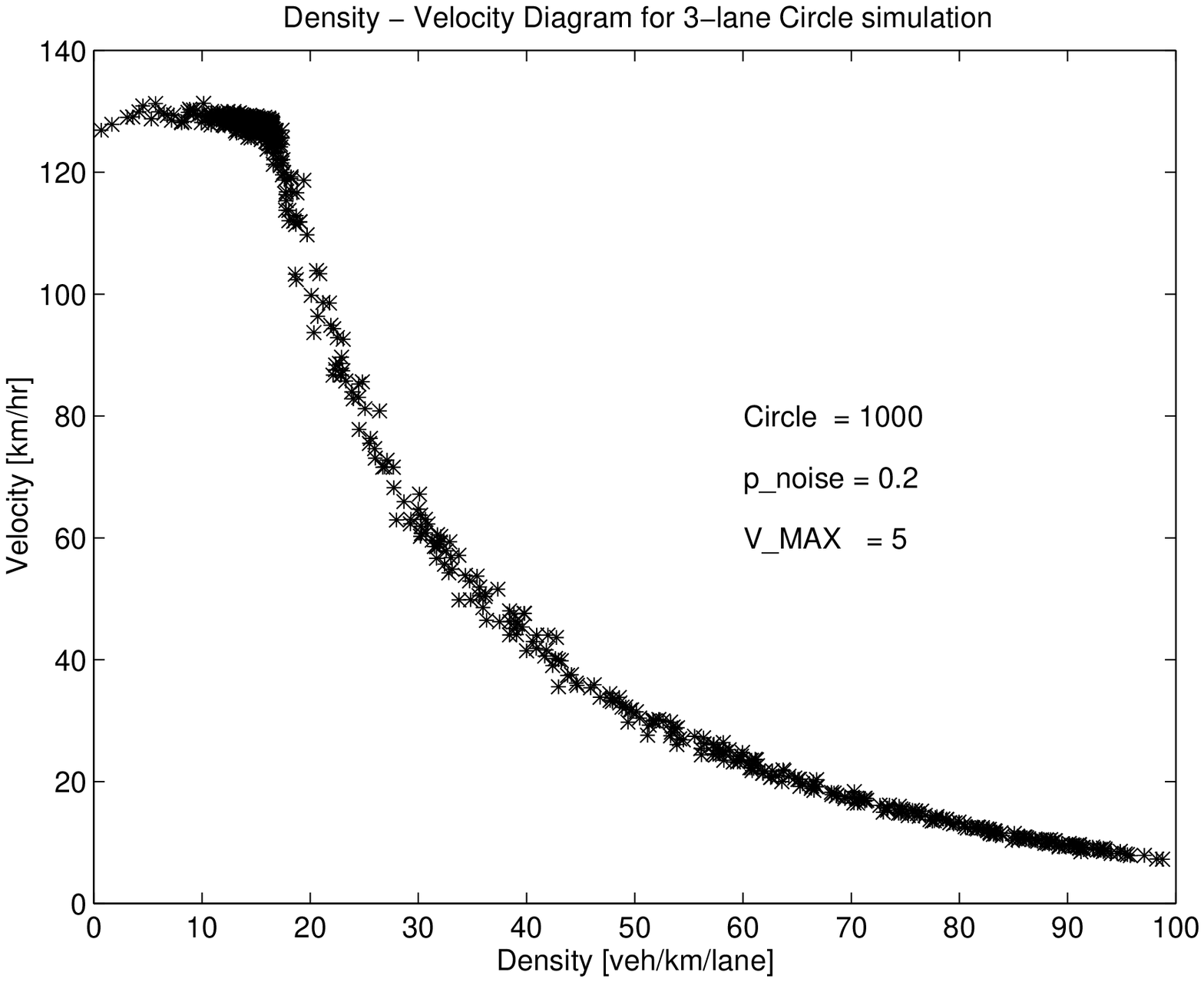}%
\hfill}
\centerline{\hfill
\epsfxsize0.5\hsize\epsfbox[47   197   553   604]{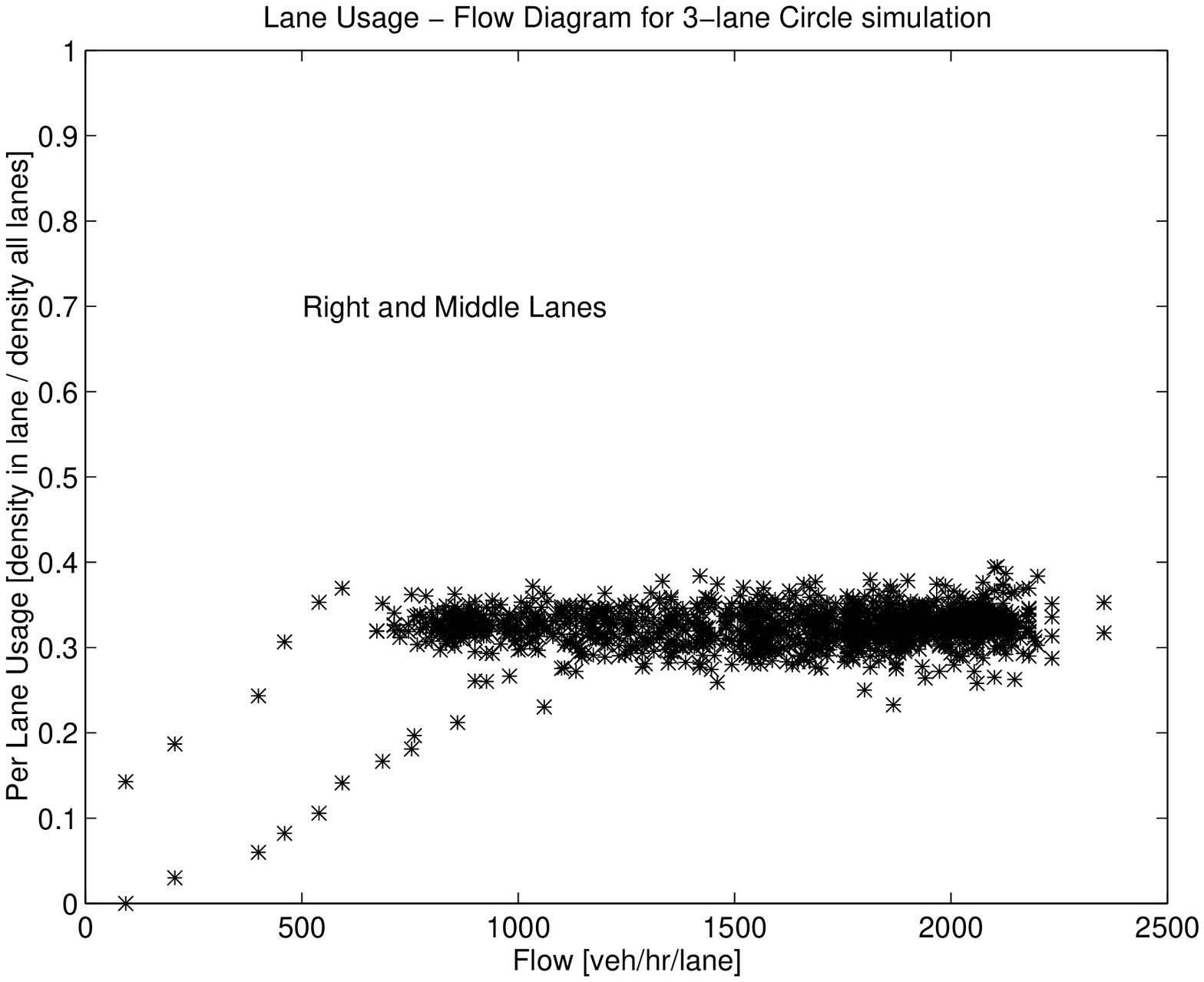}%
\epsfxsize0.5\hsize\epsfbox[47   197   553   604]{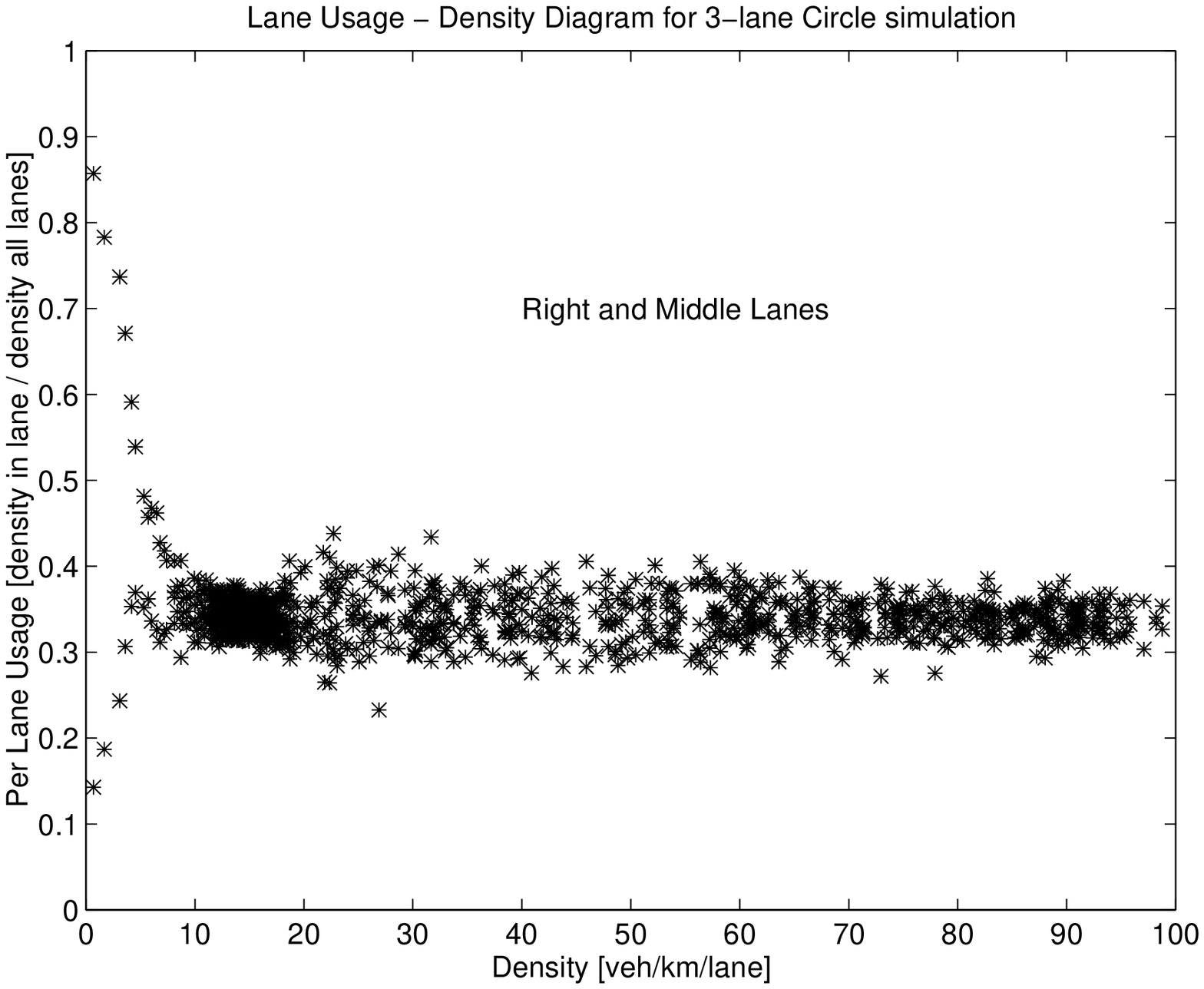}%
\hfill}
\caption{\label{3lane}%
Three-lane circle: Flow vs.\ density, travel velocity vs.\ flow, 
travel velocity vs.\ density, lane usage vs.\ flow, and land usage
vs.\ density.  The asymmetry in the lane usage at low densities is due
to the fact that the parking locations start filling in vehicles on
the right lane, and they only move to the left when traffic on the
right lane becomes dense.
}
\end{figure}

\begin{figure}
\centerline{\hfill
\epsfxsize0.5\hsize\epsfbox[47   197   553   604]{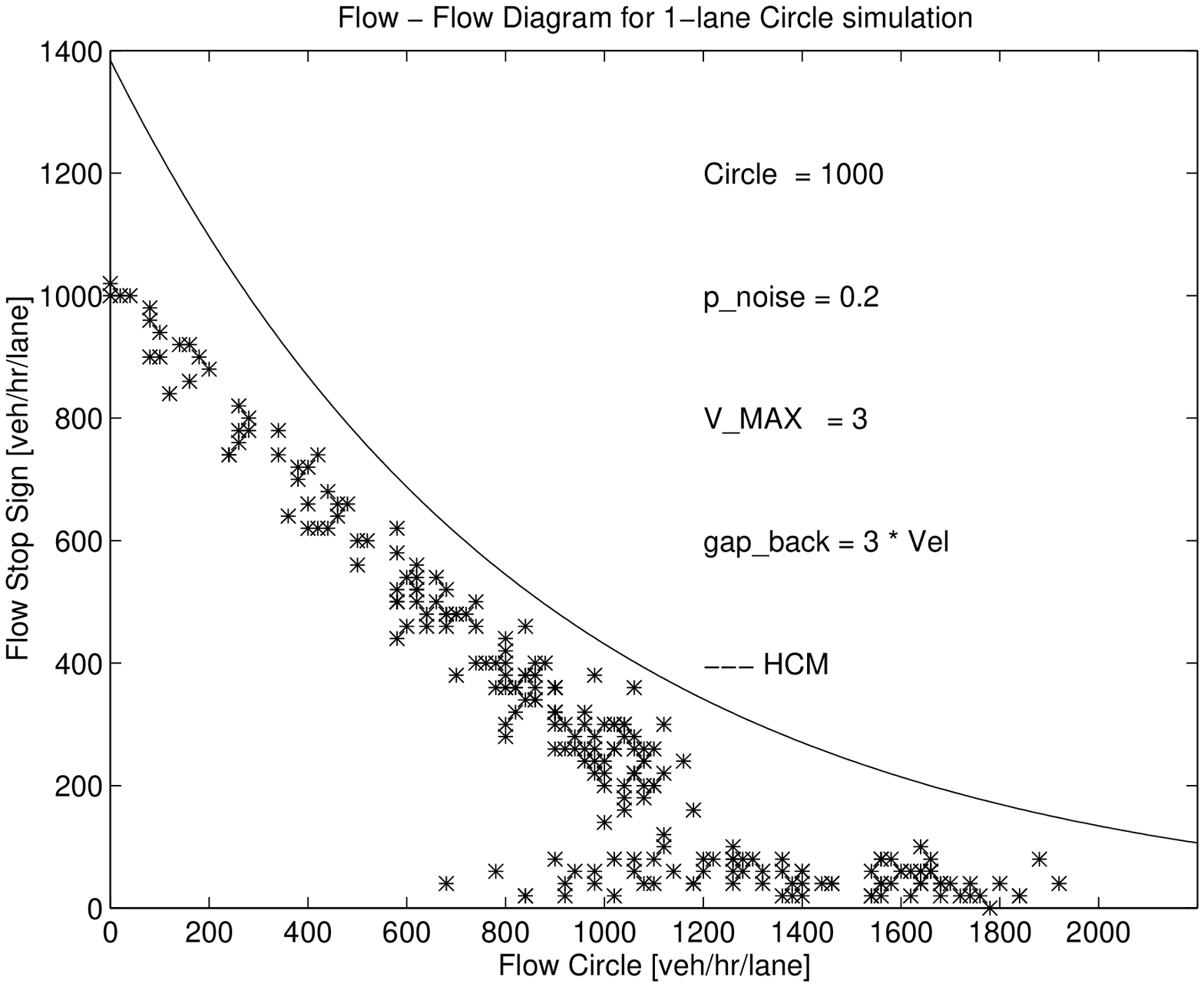}%
\epsfxsize0.5\hsize\epsfbox[47   197   553   604]{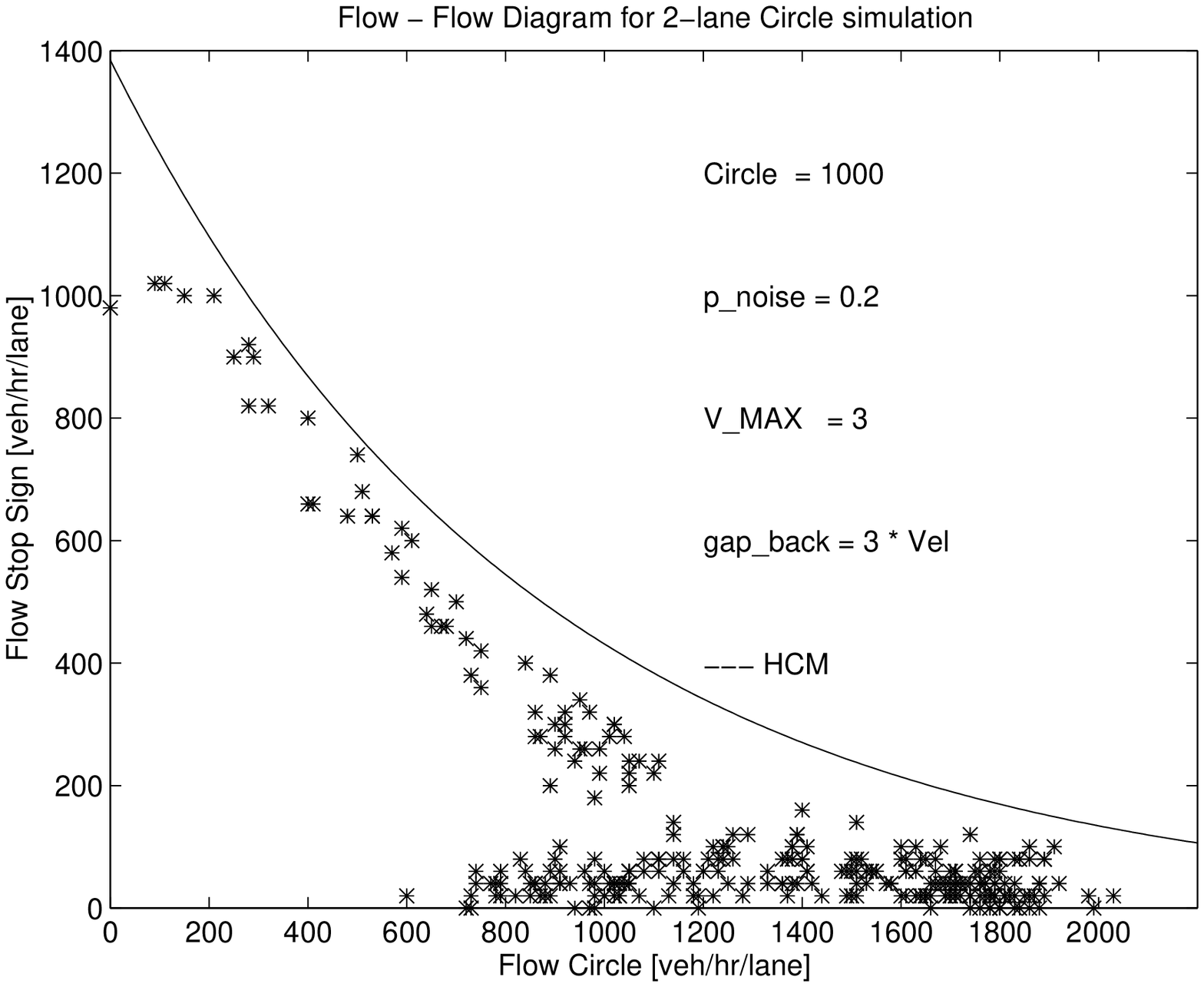}%
\hfill}
\centerline{\hfill
\epsfxsize0.5\hsize\epsfbox[47   197   553   604]{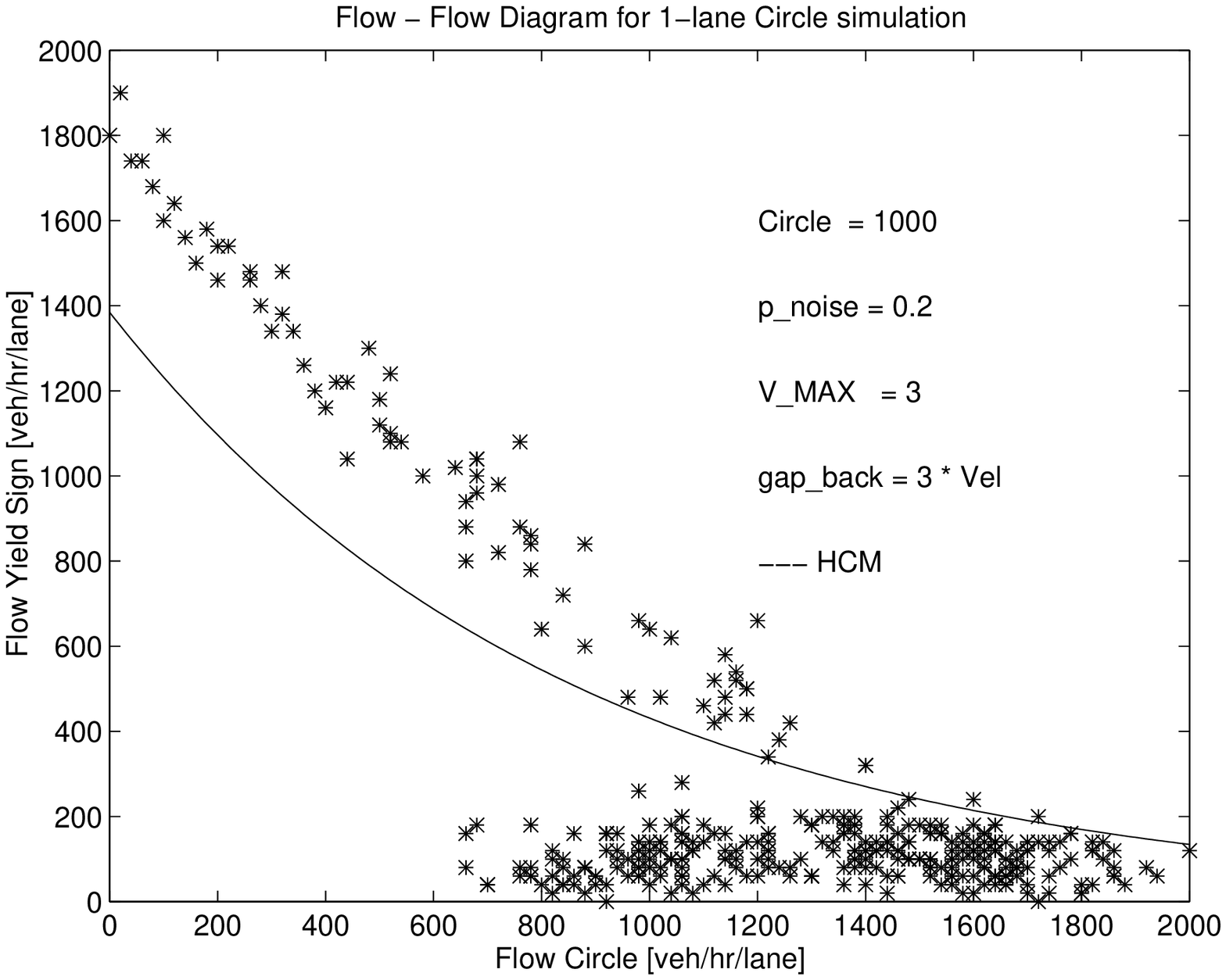}%
\epsfxsize0.5\hsize\epsfbox[47   197   553   604]{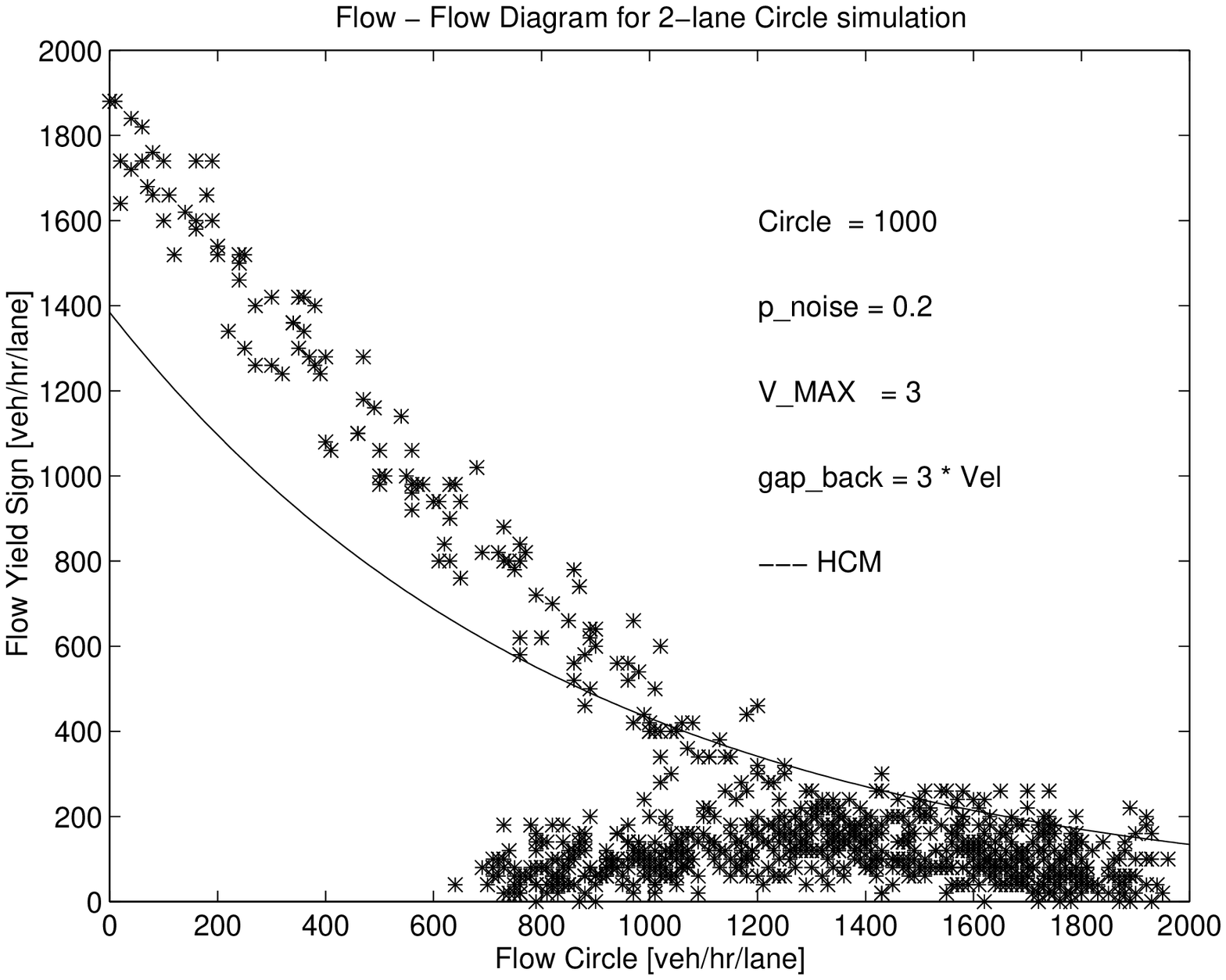}%
\hfill}
\centerline{\hfill
\epsfxsize0.5\hsize\epsfbox[47   197   553   604]{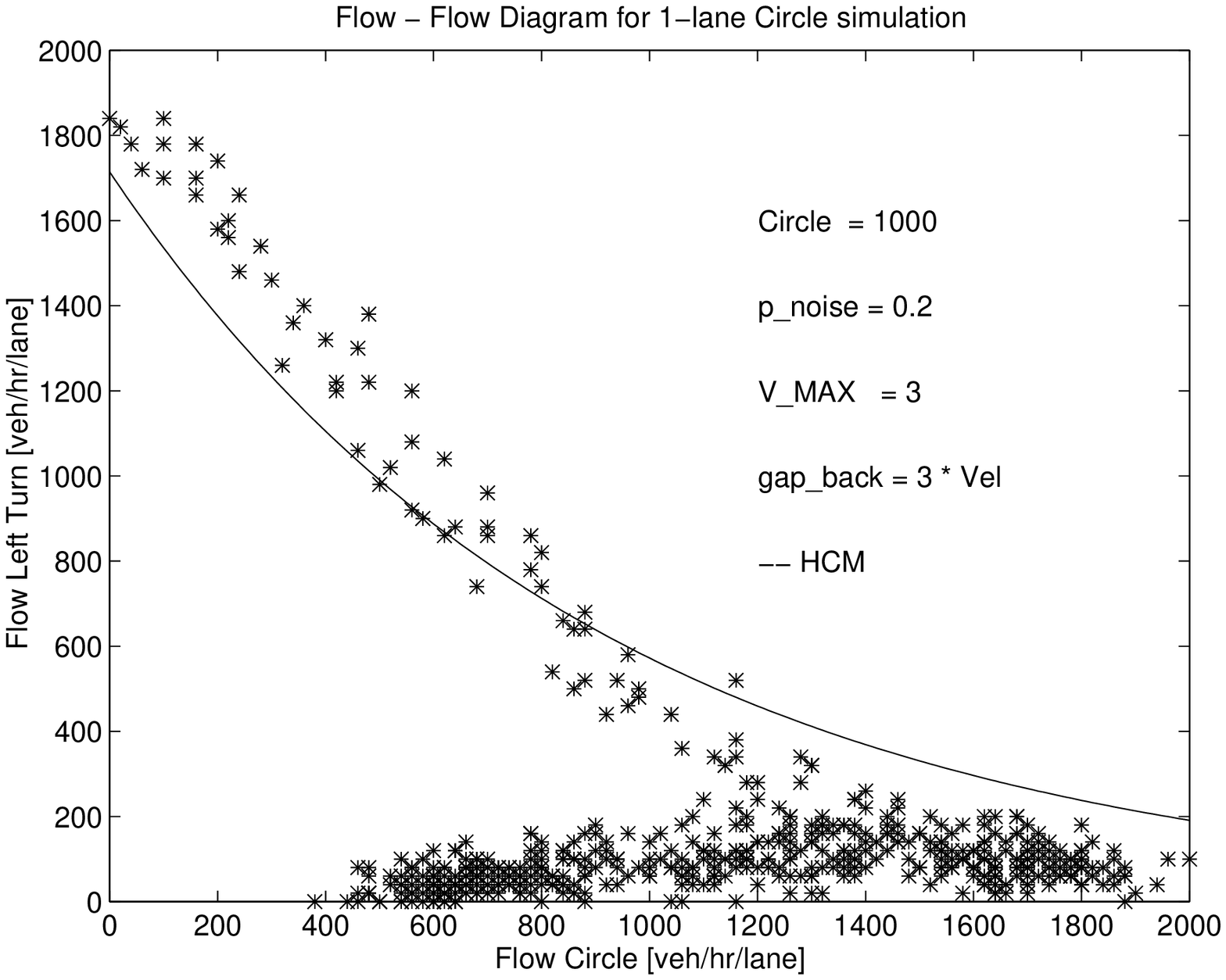}%
\epsfxsize0.5\hsize\epsfbox[47   197   553   604]{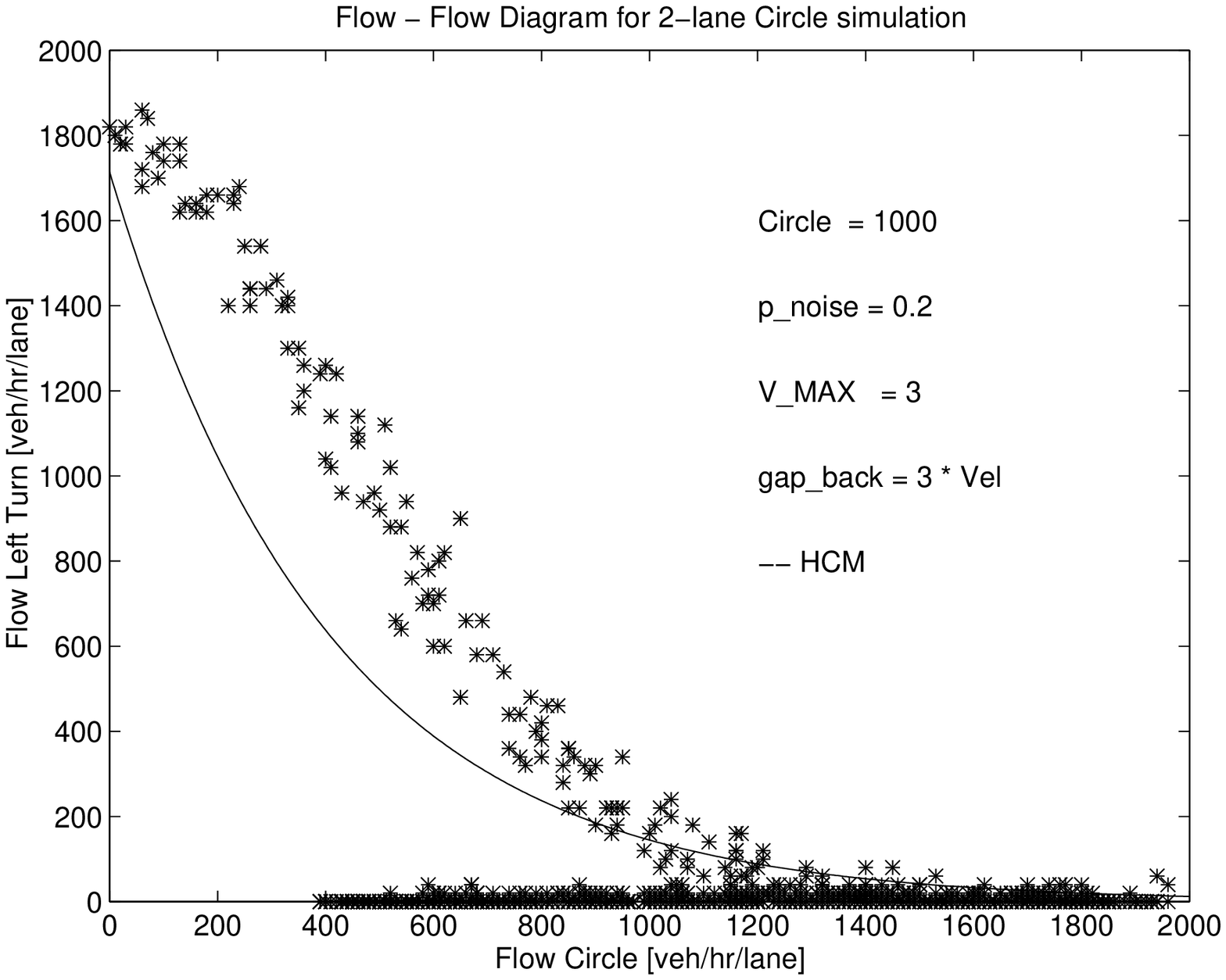}%
\hfill}
\caption{\label{unprotected-fig}%
Flow through stop sign, yield sign, and unprotected left turn.  Left
column: one-lane traffic on major road (circle).  Right column:
two-lane traffic on major road (circle).  Solid line: Highway Capacity
Manual~\protect\cite{HCM}.  Note that for ``left turn across two
lanes'' (bottom right) the interfering volume is the sum of both
lanes, i.e.\ twice the value show on the x-axis.
} \end{figure}

One problem with microscopic approaches is that, in spite of all
possible diligence, subtle differences between design and actual
implementation can make a significant difference in the macroscopic
outcome.  For that reason, this paper should also be seen as an
argument for a standardized traffic flow test suite for simulation
models.  We propose that simulation models, when used for studies,
should first run these tests to see how the macroscopic flow dynamics
actually is.  We think that the combination of results presented in
Figs.~\ref{1lane} to~\ref{unprotected-fig} are a good test set,
although extensions may be necessary in the future (e.g.\ merge lanes,
weaving, etc.).  We will attempt to provide future TRANSIMS results
also with updated versions of the results of the traffic flow tests.

\bibliographystyle{unsrt}
\bibliography{kai,ref}

\vfill\eject


\vfill\eject

\vfill\eject

\vfill\eject

\vfill\eject

\vfill\eject

\end{document}